\documentclass[conference]{IEEEtran}
\usepackage{tikz}
\usepackage{listings}
\usepackage{caption}
\usepackage{xcolor}
\usepackage{bbding}
\usepackage[linesnumbered,ruled, noend]{algorithm2e}
\let\labelindent\relax
\usepackage{enumitem}
\usepackage[utf8]{inputenc}
\usepackage{multirow}
\usepackage{array}
\usepackage{booktabs}
\usepackage{verbatim} 
\usepackage{verbatimbox}
\usepackage[tight,footnotesize]{subfigure}
\usepackage{algorithmic}
\usepackage[cmex10]{amsmath}
\usepackage{amssymb}
\usepackage{graphics}
\usepackage[hyphens]{url}
\usepackage{flushend}
\usepackage{bm}

\usepackage{letltxmacro}
\LetLtxMacro\oldttfamily\ttfamily
\DeclareRobustCommand{\ttfamily}{\oldttfamily\csname ttsize\endcsname}
\newcommand{\setttsize}[1]{\def\ttsize{#1}}%

\def\BibTeX{{\rm B\kern-.05em{\sc i\kern-.025em b}\kern-.08em
    T\kern-.1667em\lower.7ex\hbox{E}\kern-.125emX}}

\graphicspath{{pictures/}} 
\ifCLASSINFOpdf
\graphicspath{{./pictures/}}
\DeclareGraphicsExtensions{.pdf,.jpeg,.png}
\fi

\definecolor{dkgreen}{rgb}{0,0.6,0}
\definecolor{gray}{rgb}{0.5,0.5,0.5}
\definecolor{mauve}{rgb}{0.58,0,0.82}
\definecolor{groovyblue}{HTML}{0000A0}
\definecolor{groovygreen}{HTML}{008000}
\definecolor{darkgray}{rgb}{.4,.4,.4}

\lstdefinelanguage{Groovy}[]{Java}{
	keywordstyle=\color{groovyblue}\bf,
	stringstyle=\color{mauve}\ttfamily,
	keywords=[3]{each, findAll, groupBy, collect, inject, eachWithIndex, subscribe, unsubscribe},
	morekeywords={def, as, in, use, TRIGGER, CHECK, RUN, run, match, satisfy, fetch, do, foreach, branch, fetch$^{*}$},
	moredelim=[is][\textcolor{darkgray}]{\%\%}{\%\%},
	moredelim=[il][\textcolor{darkgray}]{��}
}

\newcommand{\tool}{\textsc{PFirewall}\xspace}

\newcommand{\mr}[1]{{\textcolor{black}{#1}}} 
\newcommand{\fv}[1]{{\textcolor{black}{#1}}} 

\newcommand{\zz}[1]{\textcolor{black}{#1}}

\newcommand{\Partitle}[1]{\vspace*{3pt}\noindent\textbf{#1}}

\begin{document}
\title{\tool: Semantics-Aware Customizable Data Flow Control for Smart Home Privacy Protection}



\author{
\IEEEauthorblockN{Haotian Chi}
\IEEEauthorblockA{Temple University\\
htchi@temple.edu}
\and
\IEEEauthorblockN{Qiang Zeng}
\IEEEauthorblockA{University of South Carolina\\
Zeng1@cse.sc.edu}
\and
\IEEEauthorblockN{Xiaojiang Du}
\IEEEauthorblockA{Temple University\\
dux@temple.edu}
\and
\IEEEauthorblockN{Lannan Luo}
\IEEEauthorblockA{University of South Carolina\\
lluo@cse.sc.edu}
}

\IEEEoverridecommandlockouts
\makeatletter\def\@IEEEpubidpullup{6.5\baselineskip}\makeatother
\IEEEpubid{\parbox{\columnwidth}{
    Network and Distributed Systems Security (NDSS) Symposium 2021\\
    21-24 February 2021\\
    ISBN 1-891562-66-5\\
    https://dx.doi.org/10.14722/ndss.2021.24464\\
    www.ndss-symposium.org
}
\hspace{\columnsep}\makebox[\columnwidth]{}}

\maketitle

\begin{abstract}
Internet of Things (IoT) platforms enable users to deploy home automation applications. Meanwhile, privacy issues arise as large amounts of sensitive device data flow out to IoT platforms. 
Most of the data flowing to a platform actually do not trigger automation actions, while homeowners currently have no control once devices are bound to the platform.
We present \tool, a customizable data-flow control system to enhance the privacy of IoT platform users. \tool automatically generates data-minimization policies, which only disclose minimum amount of data to fulfill automation. In addition, \tool provides interfaces for homeowners to customize individual privacy preferences by defining user-specified policies. To enforce these policies, \tool transparently intervenes and mediates the communication between IoT devices and the platform, without modifying the platform, IoT devices, or hub. Evaluation results on four real-world testbeds show that \tool reduces IoT data sent to the platform by 97\% without impairing home automation, and
effectively mitigates user-activity inference/tracking attacks and other privacy risks.
\end{abstract}

\section{Introduction}
With the prosperity of Internet of Things (IoT), smart systems 
(e.g., smart homes, factories, and hospitals)
have become realistic and are expanding with an ever-increasing speed~\cite{iot2018platforms}. 
IoT Platforms, such as Samsung SmartThings \cite{smartthings2020}, Amazon Alexa \cite{alexa2020}, openHAB \cite{openhab2019}, allow smart home users to connect heterogeneous IoT devices (e.g., sensors, actuators, appliances) and install applications on the platform to create automatic interactions among devices, i.e., home automation. 

As IoT device data flow outside,
protecting user privacy becomes critical~\cite{zheng2018user, zeng2017end}. 
Existing work protects user privacy from malicious application developers~\cite{celik18sensitive, bastys2018if, fernandes2016flowfence, tian2017smartauth, celikiotguard, nguyen2018iotsan, hsu2019safechain} or eavesdroppers~\cite{acar2018peek, datta2018developer, apthorpe2017closing, apthorpe2017spying}.
However, it is surprising that, 
while a platform receives huge amounts of privacy-sensitive data from bound IoT devices, few 
works regard the platform as untrustworthy and provide privacy protection solutions.
In fact, it is baseless to assume 
the platform is trustworthy \mr{and its data access protection is flawless, and thus we should consider 
that the rich data may be
exposed to attackers~\cite{bupadisgruntledemployee,17biggestdatabreach}.}
Furthermore, many IoT platforms share user data with partners (e.g., advertisers) 
for the expansion of businesses \cite{wink2018privacypolicy, smartthings2018privacypolicy, vera2018privacypolicy}; any improper handling may disclose privacy-sensitive data to third parties.

To support home automation, many IoT devices continuously stream data, such as sensor
events, to an IoT platform,
although most of the data actually do not trigger automation actions.
This deviates the principle of ``\emph{data minimisation}'' in European General Data 
Protection Regulation (GDPR) \cite{GDPR2016}.
We also find that no capabilities are provided for  
users to control the leakage of device data to the platform, failing to realize user-centric authorization. Therefore, \mr{we seek a privacy-enhancing system that can be used as an ``add-on'' into existing systems by privacy-conscious users. This system aims to} 1) minimize the data sent to the platform and 2) allow users to
define customizable data-protection policies for individual privacy preferences.
Multiple challenges arise for attaining these goals.

\Partitle{Challenge 1.}
Data minimization should not 
affect home automation.
We observe that the semantics of home automation apps can be represented as rules in a 
\emph{trigger-condition-action}
programming paradigm (code analysis~\cite{chi2020cross,celik2018soteria,tian2017smartauth} and natural language processing~\cite{tian2017smartauth, zhang2018homonit, ding2018safety} have proven effective in extracting rules from apps). Our insight is that, according to the rule semantics, 
we could reduce data leakage without impairing automation. For example, given a rule ``\emph{when a motion is detected (\mr{\emph{\textbf{trigger}}}), if the indoor temperature is higher than 79$^{\circ}F$ (\mr{\emph{\textbf{condition}}}), turn on the A/C (\mr{\emph{\textbf{action}}})},''
we could derive data-minimization policies, 
such that
1) if the temperature is not higher than 79$^{\circ}F$,
no motion or temperature data are sent to the platform; 2) if the A/C is already on (that is, the rule execution will not
change anything), 
no data need to be sent, 
and 3) otherwise, the temperature and the motion data must be sent, but the temperature value can be obfuscated to a random value larger than 79$^{\circ}F$.

\Partitle{Challenge 2.}
Many platforms are closed systems that do not allow platform-level modifications, and
it is probably unrealistic to expect a platform to cooperate to enforce 
data minimization. Thus, how to enforce data-protection policies in closed systems is a challenge. One may propose to \emph{avoid} this challenge
by building a new purely-local platform, such that no data have to flow out of a home, or one can simply cut the network cable of a local gateway~\cite{mozillaiot2019} and enforce the home automation locally.
However, most of the leading platforms (IFTTT, SmartThings, Google Home, Amazon Alexa) employ a cloud-based architecture due to its advantages in many aspects such as storage, integrated services, management. Thus, existing users may not be willing to give up their choice of platforms.


\mr{Inspired by the concept of firewalls \cite{coley1998firewall, wesinger1999firewall}}, we design \tool as a mediator,
which sits between IoT devices and the hub (or the platform backend cloud) to transparently filter data based on
privacy-protection policies. \mr{However, our \tool is significantly different from the traditional firewalls. First, the “trigger-condition-action” paradigm of IoT automation rules is different from that of traditional firewall rules. Second, PFirewall utilizes application-layer data and performs semantics-aware data filtering, while traditional firewalls do not use application-layer payload data.}
There is another challenge for implementing \tool:
the original communication between IoT devices and the hub (or the platform backend cloud) is encrypted,
which prevents \tool from understanding it and then filtering data.
We overcome this challenge with a %
\emph{virtualization}
approach:
On one hand, the data mediator acts as a hub to pair
with all the IoT devices. On the other hand, \tool creates a virtual device (for each real IoT device), which connects with the hub (or platform backend cloud). 
This way, neither the IoT devices nor the platform needs to be modified.


We demonstrate the ideas by implementing \tool on three representative IoT platforms: \emph{SmartThings classic}, \emph{the new SmartThings}\footnote{\fv{SmartThings supported the two major versions at the time of research. Although the SmartThings classic mobile app was discontinued in October 2020, most features concerned in this paper such as SmartApps are now included in the new system.}} and \emph{openHAB}, where the former two are \emph{cloud-based} IoT platforms and the third a \emph{gateway-based} platform. We evaluate \tool in four real-world testbeds. The results show that \tool reduces the amount of data sent to the platform by 97\%, without affecting  home automation.  Our performance evaluation also shows that the data reduction severely impairs the attacker's ability to infer and track privacy-sensitive behaviors, such as bathroom usage, home occupancy, etc. An earlier version of this paper was posted on \emph{arXiv} in October 2019~\cite{chi2019pfirewall}.

The contributions of this work are summarized as follows.
 
\begin{itemize}[leftmargin=*]
	

	\item We design an effective data flow control system to protect user privacy in home automation. Data-minimization policies are automatically generated based on automation apps, reporting only the needed data for app execution (minimizing data out). Moreover, we provide users an easy-to-use tool to define and customize their own privacy policies (e.g., \emph{``during the sleep mode, no data should be sent out from devices in bedrooms''}). 
	
	\item We overcome the challenges that most IoT platforms and IoT devices are closed-systems and cannot be modified to support the proposed data filtering, and implement our solution on the standard wireless communication protocols including ZigBee, Z-Wave and WiFi.

	\item We implement a prototype to work with various IoT devices and three popular platforms: SmartThings classic, the new SmartThings and openHAB. Through the evaluation in four real-world testbeds, we demonstrate that our system significantly reduces the privacy risks due to data leakage and it causes very small latency to home automation.
\end{itemize}

\section{Background: Smart Home Platforms}
\label{section_background}
\begin{figure}[t] 
	\centering
	\includegraphics[width=0.45\textwidth]{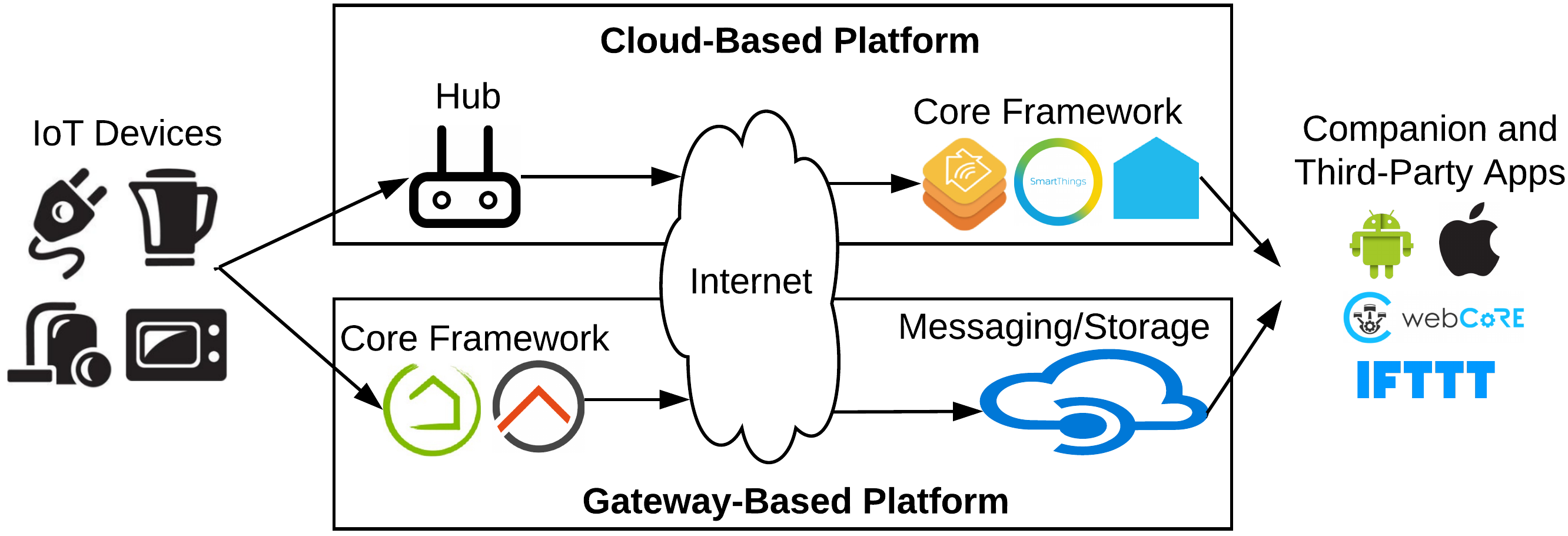}\\ 
	\caption{Smart home platform architecture.}\label{fig_smartthings} 
 	\vspace{-10pt}
\end{figure}

Smart home platforms can be categorized into cloud-based platforms (CBPs) and gateway/hub-based platforms (GBPs), according to whether the core framework of a platform is hosted in a remote cloud or a gateway/hub device located at home (as shown in Fig.~\ref{fig_smartthings}). 
Note that the gateway running a
core framework at home does not resolve the privacy leakage threats completely, as the
gateway connects to the Internet and is under the full control of the platform cloud administrator. 
We choose a CBP---\emph{SmartThings}, one of the most popular and full-fledged 
platforms, as an example to describe the key components in a smart home system.

\begin{itemize}[leftmargin=*]
\item \textbf{Hub.}
A CBP hub connects IoT devices through distinct short/medium-range wireless radios (ZigBee, Z-Wave, etc.). 
The hub plays a key role to ensure the interconnectivity and interoperability of heterogeneous IoT devices. A GBP also has a hub-like device\footnote{We use \emph{gateway} for distinguishment.} which not only connects IoT devices but also hosts the core framework (described below). Note that the hub or gateway device, though physically located at home, is conceptually regarded as a part of the platform in terms of data privacy protection in that it is under the full control of the platform cloud administrator.

\item \textbf{Cloud.}
The backend cloud of a CBP hosts the core framework and provides cloud messaging, storage and other necessary services. The cloud in a GBP is typically responsible for messaging and storage. The cloud messaging service facilitates some critical functionalities, such as notification, third-party application integration, remote monitoring and control.
Many Internet-based services depend on the cloud. 

\item \textbf{Core Framework.}
The core framework runs major functionalities of a platform, including home automation. SmartThings classic provides a sandboxed runtime environment for running \emph{device handlers} and \emph{SmartApps}. Device handlers are software wrappers that abstract the physical devices
as a set of \emph{capabilities} and they handle the underlying protocol-specific communications between the core framework and the physical devices. \mr{
The core framework represents each physical device as a \emph{device instance} by instantiating the corresponding device handler.}
Automation rules can be defined by installing SmartApps or configuring the rule-creating interfaces on the companion mobile app.

\item \textbf{Companion and Third-Party Apps.}
To provide a convenient user interface (UI) for users to manage their hubs, IoT devices 
and apps, a platform usually provides a smartphone \emph{companion} app. 
For instance, in the SmartThings companion app, users can install and configure a SmartApp.

\end{itemize}

\section{Motivation and Threat Model}
In this section, 
\mr{we first discuss two privacy concerns that motivate this work,} 
and then present the threat model.

\subsection{Privacy Concerns about Platforms}
\label{subsec_privacy_implications}
\subsubsection{Trust By Default}
In smart home systems, the platforms are typically fully trusted. After being installed, a platform gains the access privilege to all connected home IoT devices technically by design and legally, by claiming a \emph{terms and conditions} or a \emph{privacy policy}. 
\fv{After that, IoT platforms receive rich data from connected devices, 
no matter whether the data are required or not for providing services. 
To demonstrate this fact, 
we conduct an experiment on an exemplar platform--SmartThings.}


\vspace{3pt}
\mr{\emph{Are IoT data continuously flowing out of homes?}} 
To answer this question, we connected four ZigBee devices (a multipurpose sensor, a motion sensor, an arrival sensor and a smart outlet) and a Z-Wave sensor (Aeotec Multisensor 6) to a SmartThings hub and observe the data received by the SmartThings cloud on its logging interface \cite{smartthings2020ide}. We did \emph{not} install any automation apps and did \emph{not} operate any SmartThings-provided interfaces. We only interacted with the IoT devices physically. We found that the platform cloud kept receiving device data (e.g., motion, switch, temperature, etc.) from devices, indicating that device data flow out via the hub \emph{even when they are not needed by any automation}. \fv{Besides our experiment, traffic analysis researches \cite{trimananda2020packet, oconnor2019blinded} also show that certain traffic patterns for transmitting device events from the SmartThings hub to the backend cloud can be observed when the corresponding events are generated by devices, which is consistent with our result.} 

\fv{Reasons for this fact include: (1) to enhance interoperability, IoT devices are designed to
simply send out all data to the paired hub or cloud for further processing, but unaware of how the data will be consumed; (2) most platforms support more than one service (e.g., home automation, dashboards for viewing and changing device states), and therefore do not deploy a data filtering component on the hub devices (manufactured by them) or at the front-end to filter out data that are not needed for specific services. However, we argue that the system could be improved or patched to enhance user privacy in different services. In this paper, we focus on enhancing user privacy in the home automation service, which is context-aware and more difficult to address than other services. We also discuss a possible solution for the dashboard service in Section~\ref{section_discussion_limitations}.}


\subsubsection{Limited User Capabilities}
Currently, users have few (or no) capabilities to 
control device data sent to a platform~\cite{alrawi2019sok}. They only have a binary choice: either connect a device to the platform or not; once connected, the device continuously reports data to the platform.

\subsection{Threat Model}
We consider that a smart home platform may be exploited by attackers for accessing user private data and inferring user privacy-sensitive behaviors. \mr{For example, an attacker may find a way to gain access to the data; also the platform administrators may be curious to mine user privacy.}
\mr{We do not assume that the platform is attacker-controlled, but aim to handle the concern that the rich data 
may be exposed to attackers and exploited for malicious purposes.}
The platform is assumed to be honest, i.e., it faithfully operates all services and does not perform active attacks on users.
Attacks that exploit vulnerabilities of home IoT devices ~\cite{wurm2016security, acar2018web}, side channels~\cite{acar2018peek, apthorpe2017spying, ronen2016extended}, or home local networks~\cite{chomsiri2008sniffing, arbaugh2002your} to steal private data are not within the scope of this work. 

Malicious IoT apps 
may request more device data than what is needed by the claimed functionality in the app descriptions. Detecting and preventing such malicious apps have been well studied, e.g.,~\cite{fernandes2016security, tian2017smartauth}. In this work, we assume that this kind of malicious apps have been inspected by an existing solution such as \cite{tian2017smartauth}.
Moreover, many platforms (e.g., SmartThings, Amazon Alexa) allow users to create rules on mobile app interfaces, largely eliminating such attacks. 

\tool itself contains a firewall that only allows traffic from whitelisted sources (attacks targeting \tool are further discussed in Section~\ref{section_discussion_limitations}).

\section{System Overview}
To protect data privacy, IoT data are processed before they leave a smart home. 
In this work, we propose a privacy-preserving data minimization approach that can guarantee the correctness and completeness of the desired home automation, satisfy personal privacy preferences, and significantly reduce the amount of data sent to the platform. In order to achieve these goals, three challenges need to be addressed.

First, data protection must not accidentally affect the correctness and/or completeness of
home automation (Section~\ref{section_ap}). Second, it is challenging to enforce the data protection without modifying the IoT devices, hub, or platform framework \mr{which are usually closed systems} (Section~\ref{section_relay}). Third, it is non-trivial to provide user-friendly interfaces for non-expert users to define their own privacy-protection policies  (Section~\ref{section_up}).

Most IoT devices can be categorized as the following two cases
(Section~\ref{section_device_connector} presents our survey of commercial IoT devices): (1) IoT devices use open-source wireless technologies (e.g., ZigBee, Z-Wave, and Bluetooth) to connect with an in-home hub, and \mr{(2) IoT devices use WiFi to connect, in an end-to-end encryption manner, with their \emph{vendor cloud}, which then bridges IoT devices to IoT platforms through encrypted channels.} In case (1), \tool can act as the hub to directly connect with IoT devices, and hence can see the packet contents from the devices. \mr{In case (2), \tool cannot act as a hub to directly talk with these devices but instead could access the devices via the REST (REpresentational State Transfer \cite{restapi}) APIs provided by the vendor cloud.} With the above observations, we propose to place a \emph{mediator}
between IoT devices and the platform to intervene in the communication between them. The mediator makes it possible to process the \emph{raw} IoT data before forwarding them to the platform. With this insight, we design \tool, a system that can enforce strong privacy protection and user privacy preference, while not affecting the home automation at all. 

\begin{figure}[tb]
	\centering
	\includegraphics[width=0.49\textwidth]{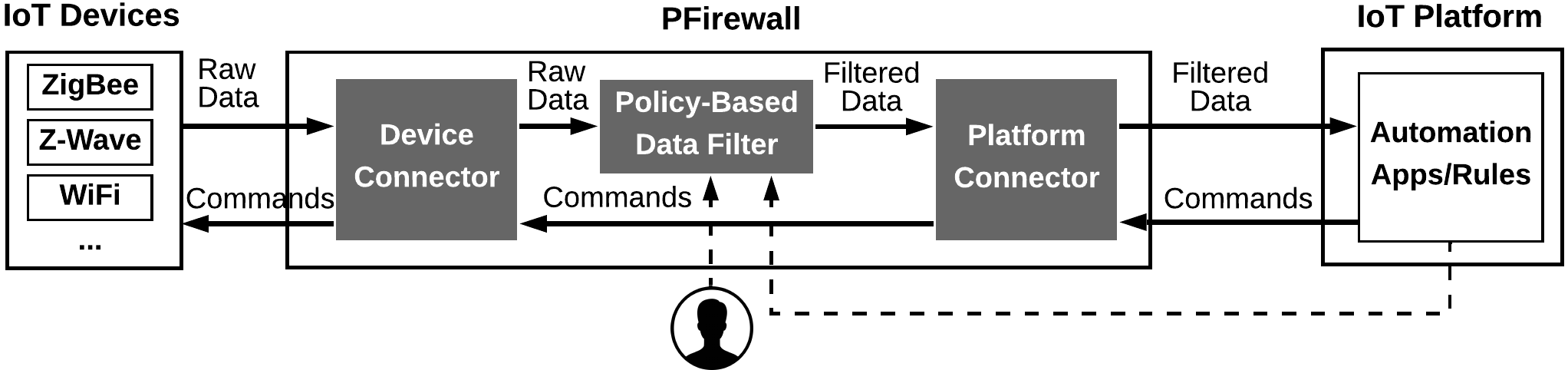}\\
	\caption{\mr{Architecture of \tool}. Solid arrows: device data\mr{/command} flows; dotted arrows: inputs for generating policies. \tool can automatically generate policies from apps, so user-specified policies are \textbf{\emph{optional}}.}\label{fig_architecture}
 	\vspace{-10pt}
\end{figure}

As shown in Fig.~\ref{fig_architecture}, \tool comprises these modules:

\begin{itemize}[leftmargin=*]
	\item \mr{\textbf{Device Connector}} talks with IoT devices directly via device-dependent protocols such as ZigBee, Z-Wave, or indirectly via vendor cloud APIs (see Section~\ref{section_device_connector}). It also forwards the raw device data to the \emph{policy-based data filter} for processing.
	
	\item \mr{\textbf{Platform Connector }}
	interacts with the platform through specific connectivity technology supported by the platform, on behalf of the physical devices connected to the \emph{device connector} (Section~\ref{section:connect_platform}). The \emph{device connector} and \emph{platform connector} collaboratively mediate data between IoT devices and the platform. 
	
	\item \textbf{Policy-Based Data Filter} sits between the device connector and platform connector to filter the sensitive data (i.e., events) from IoT devices based on policies. It has three main components: \emph{policy generator}, \emph{conflict detector} and \emph{policy engine}. The \emph{policy generator} generates data-filtering policies in two ways: (1) it takes automation rules as input to generate data-minimization policies (Section~\ref{section_ap}) and (2) it transforms user-specified policies (from the user interfaces) into executable-formatted policies (Section~\ref{section_up}). 
	\emph{Conflict detector} inspects if a user-specified policy conflicts with existing data-minimization policies and reports conflicts to users (Section~\ref{section_policy_conflict}). \emph{Policy engine} interprets and executes all policies. 
	
\end{itemize}

\mr{Note that while \tool filters device events before reporting them to the platform, it does not filter commands, i.e., all commands received from the platform by the platform connector are forwarded to the physical devices via the device connector.}

\section{Design and Implementation}
\label{section_design_implementation}
In this section, we present the design and implementation details. 
We first describe the policy-based data filter for contextually controlling IoT data flows (Section~\ref{section_policy_system}). Then, we present the mediator for enforcing policies in existing IoT systems (Section~\ref{section_relay}). To demonstrate the applicability of \tool, we show how \tool integrates with three popular platforms: SmartThings classic, the new SmartThings and openHAB (Section~\ref{section:connect_platform}).

\subsection{Data Filtering Policies}
\label{section_policy_system}
\tool filters data based on two types of policies: automation-dependent data-minimization policies (APs) and user-specified policies (UPs). To achieve data minimization, i.e., only reporting the minimum amount of data that are necessary for home automation, automation rules are analyzed to generate the corresponding APs. UPs are generated from a user interface provided by \tool and work with APs simultaneously, which is an important supplement to customize privacy preferences that cannot be learned from home automation.
\subsubsection{Automation-Dependent Data-Minimization Policies}
\label{section_ap}
\mr{Home automation systems run user-customized \emph{trigger-condition-action} rules, each of which can be expressed as
``\emph{when [trigger], if [condition], then [action]}''~\cite{bae2004automatic, brackenbury2019users}.}
Each rule runs reactively, i.e., it reacts when its \emph{trigger} is satisfied by a new event (termed as \emph{trigger event}) and then executes \emph{action} if the smart home is under the prescribed \emph{condition}. Note that the trigger and condition are different: a trigger describes a constraint that checks against the new events, while a condition includes a set of constraints that check on the static states (e.g., current device states, time, etc.).
An example rule $\mathbf{R}_{1}$ is 
``\emph{when a presence sensor $ps_{1}$ becomes present (trigger), if the reading of a temperature sensor $ts_{1}$ is higher than 86$^{\circ }F$ (condition), then turn on the fan $f_{1}$ (action)}.'' 
\mr{Some automation rules follow \emph{trigger-action}, i.e., ``\emph{when [trigger], then [action]}'', which is a special case of the \emph{trigger-condition-action} paradigm where \emph{condition} is always true.\footnote{To avoid confusion, we explicitly use \emph{when} and \emph{if} to distinguish trigger and condition, respectively.} For example, a trigger-action rule $\mathbf{R}_{2}$ is defined as ``\emph{when the reading of a temperature sensor $ts_{1}$ is higher than 86$^{\circ }F$, then turn on the fan $f_{1}$''}.} Without loss of generality, we will use the rule $\mathbf{R}_{1}$ as a running example to present how \tool filters data.

\vspace{4pt}
\noindent\textbf{Automation Rule Extraction.}\vspace{2pt}
Extracting automation rules from rule-creating interfaces is the first step for AP generation. Automation rules are generated by installing IoT apps or using rule templates on web/mobile app interfaces. The rule extraction regarding both methods has been widely studied by state-of-the-art literature. Code analysis has been shown to be an effective way to extract rule semantics from IoT apps. For example, by utilizing Abstract Syntax Tree (AST) analysis on SmartApps, \cite{fernandes2017internet} identifies requested and used capabilities in SmartApps, \cite{tian2017smartauth, ding2018safety} break down SmartApps and extract rule information, \cite{jia2017contexiot, zhang2018homonit, celikiotguard} build Deterministic Finite Automatons (DFAs) from SmartApps. Symbolic execution is a powerful technique to analyze rule semantics from apps \cite{chi2020cross, celik2018soteria}. Text data crawling and natural language processing have been used for rule extraction from mobile apps and web pages \cite{zhang2018homonit, hwang2016data}.
Rather than develop new tools, in this paper, we adapt the solution provided in \cite{chi2020cross} to extract rules from SmartThings classic and manually encode rules defined in the new SmartThings and openHAB for which we are unable to find an open-source implementation.
We envision that more rule extraction tools will be developed and made publicly available, which can eliminate the manual efforts for encoding rules.

\vspace{4pt}
\noindent\textbf{Policy Generation.}\vspace{2pt} 
Consider the example rule $\mathbf{R}_{1}$.
By default, the event streams from devices (presence sensor, temperature sensor, fan) are reported to the platform for executing $\mathbf{R}_{1}$.  
However, we observe that these data are not all required for executing $\mathbf{R}_{1}$ in cases:

\begin{enumerate}[leftmargin=*]
\renewcommand\labelenumi{(\theenumi)}
    \item The presence sensor $ps_{1}$ does not send any event;
    \item $ps_{1}$ sends a ``not present'' event; 
    \item The temperature measured by $ts_{1}$ is lower than 86$^{\circ }F$;
    \item The fan $f_{1}$ is ``ON'';
    \item $ps_{1}$ sends a ``present'' event and the last reported temperature by $ts_{1}$ is higher than 86$^{\circ }F$. 
\end{enumerate}

In cases (1)-(4), there is no need to report any data from $ps_{1}$ and $ts_{1}$, as well as data from other devices if they are not used in other rules. In case (5), it is unnecessary to report the temperature data since the temperature value stored in the platform database \mr{(i.e., the last reported value)} satisfies the rule condition checking. In no cases, the ON/OFF state of $f_{1}$ is useful for executing $\mathbf{R}_{1}$. In addition, we can report a random temperature value that is higher than 86$^{\circ }F$ instead of the real one when reporting a temperature value is necessary for rule execution. From this example, we can see that only a small portion of the device data are required for home automation, which motivates us to run data-minimization policies that can significantly reduce the amount of data going out while not affecting any automation rules. \mr{Given the unique trigger-condition-action paradigm, the existing access control policy parsers cannot be directly used. Therefore, we develop a customized context-aware policy format and a corresponding policy engine for the trigger-condition-action paradigm.}

\begin{lstlisting}[float,floatplacement=t,caption={Context-aware policy format},label=listing_policy,captionpos=b, abovecaptionskip=5pt, belowcaptionskip=15pt, numbers=right]
TRIGGER:{
	match (:type).(:subject).(:attribute)
	satisfy (:operator)->(:value)
	[fetch*] (:type).(:subject).(:attribute*)
	[branch] (:operator1)->(:value)
	run (:method)(:paras)(:delay)
	[else] (:method1)(:paras1)(:delay) }
CHECK: [{
    fetch (:type).(:subject).(:attribute)
	satisfy (:operator)->(:value)
	[fetch*] (:type).(:subject).(:attribute*)
	[branch] (:operator)->(:value)
	[run] (:method)(:paras)
	[else] (:method1)(:paras1) }, ...]
\end{lstlisting}

\begin{figure}[tb]
	\centering
	\includegraphics[width=0.41\textwidth]{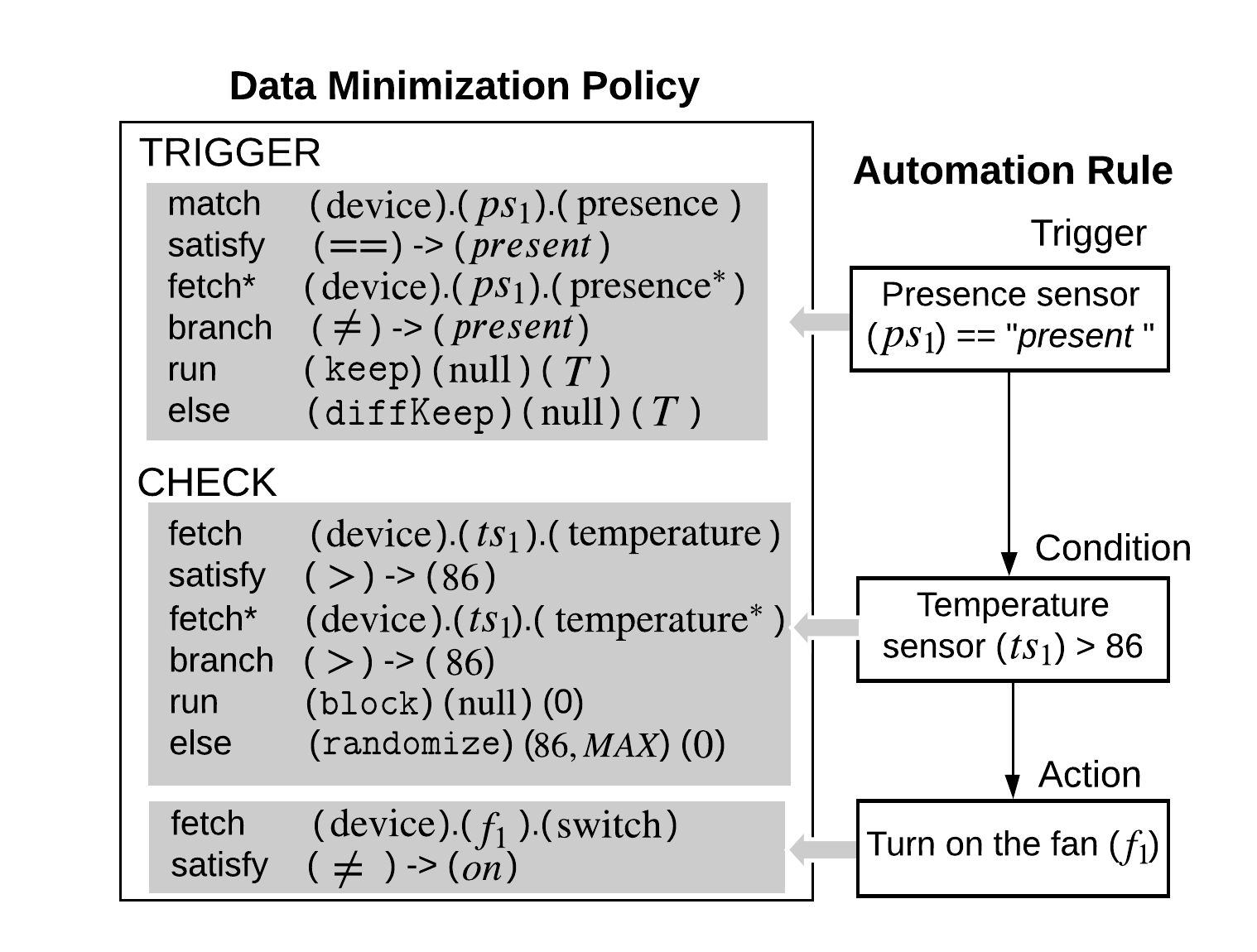}\\
	\caption{Policy derivation from an automation rule.}\label{fig_policy_from_rule}
	\vspace{-10pt}
\end{figure}

Formally, we define a data flow policy as $\mathbf P\tt \mathbf=(T, C)$, where $\tt T$ and $\tt C$ denote the \texttt{TRIGGER} and \texttt{CHECK} section in a policy, respectively, as shown in Listing \ref{listing_policy}. \mr{Fig.~\ref{fig_policy_from_rule} illustrates an policy instance that is derived from the running rule example $\mathbf{R}_{1}$.} \texttt{TRIGGER} defines the incoming event that awakens the execution of $\mathbf P$ and \texttt{CHECK} encapsulates a list of items, each of which contains a constraint that must be satisfied for the policy to indeed perform actions. \texttt{TRIGGER} is derived from the \texttt{Trigger} of an automation rule (e.g., $\mathbf{R}_{1}$) and \texttt{CHECK} is derived from the \texttt{Condition} and \texttt{Action} of the same rule.

\texttt{type} indicates that a datum is related to a device or to time; \texttt{subject} is used to identify a specific IoT device (i.e., device ID); {\tt attribute} specifies the attribute of a device (a device may have multiple attributes) or a time-related feature (e.g., time of day, date, timer). Fields \texttt{type}, \texttt{subject} and \texttt{attribute} are used to uniquely identify an event type \mr{(e.g., an event of a device $ps_{1}$'s presence)} or a state type \mr{(e.g., the reading of a device $ts_{1}$'s temperature)}. Hence, these fields in \texttt{TRIGGER} indicate that only data of the specific event type \mr{(e.g., $ps_{1}$'s presence)} trigger the execution of the policy $\mathbf P$; similarly, these fields in \texttt{CHECK} indicate what states \mr{(e.g., $ts_{1}$'s temperature and $f_{1}$'s switch)} $\mathbf P$ should query for constraint satisfaction checking. Fields \texttt{operator} and \texttt{value} define a constraint to the event \mr{($ps_{1}$'s presence is ``present'')} or state \mr{($ts_{1}$'s temperature is higher than 86)} in the same block. 
The policy $\mathbf P$ will take actions only when all the constraints defined in \texttt{TRIGGER} and \texttt{CHECK} are satisfied; if any constraint is not satisfied, the policy execution aborts and will not perform actions. \mr{Recall the aforementioned cases (1)-(5). The policy in Fig.~\ref{fig_policy_from_rule} only needs to report data for ensuring the correct execution of $\mathbf{R}_{1}$ when a new ``present'' event is observed from $ps_{1}$ and at the same time $ts_{1}$'s current temperature is higher than 86 and $f_{1}$ is off.}

\begin{table}
	\scriptsize
	\caption{Summary of methods used in data flow policies}
	\renewcommand\arraystretch{1}
	\newcommand{\tabincell}[2]{\begin{tabular}{@{}#1@{}}#2\end{tabular}}
	\newcolumntype{P}[1]{>{\arraybackslash}p{#1}}
	\newcolumntype{M}[1]{>{\centering\arraybackslash}m{#1}}
	\begin{tabular}{P{2.8cm}P{5cm}}		
		\toprule
		\bf Method & \bf Description\\ \midrule
		\texttt{keep()} & Report the original value\\ 
		\texttt{block()} & Do not report\\
		\texttt{diffKeep(v)} & Report a different value ($\neq$\texttt{v}) and then value \texttt{v}\\ 
		\texttt{randomize(v1,v2)} & Report a random value $\in$ (\texttt{v1},\texttt{v2})\\ 
		\texttt{startTimer(id)} & Create or reset a timer with identity \texttt{id} \\
        \texttt{stopTimer(id)} & Stop and reset a timer with identity \texttt{id} \\
        \texttt{fireTimer(id)} & Fire a timer \texttt{id} and execute actions in its callbacks \\
        \texttt{addCallback(id,act)} & Add an action \texttt{act} to the callbacks of timer \texttt{id} \\
		\bottomrule 
	\end{tabular}
	\label{table_methods}
\end{table}

Policy actions defined in \texttt{run} or \texttt{else} indicate how to report data to the platform. \texttt{method} and \texttt{parameters} define how to process the corresponding raw data and \texttt{delay} controls the timing for reporting the processed data to the platform.
Table~\ref{table_methods} shows a summary of all the supported methods. In the default setting, binary-value sensors such as the presence sensor reports binary values alternatively; thus, the platform only fires an event when observing a value change. Our data filtering may affect the alternation of ``present'' and ``not present'' values in the data stream of $ps_{1}$. 
When the platform receives ``present'' but finds the last received value is also ``present,'' it will not fire a new ``present'' event in its framework and thus $\mathbf{R}_{1}$ cannot be triggered. The policy uses \texttt{diffKeep()} to address this issue; \texttt{diffKeep()} reports ``not present'' followed by ``present'' with a time delay $T$, which ensures a ``present'' event is fired on the platform\footnote{We fine-tuned the value of $T$ and found a sweet spot as small as 300 millisecond without causing failure in the three platforms.}. However, if the last reported value is ``not present'', the policy only needs to use the method \texttt{keep()} which reports the received value ``present''.
To handle the above situations in \texttt{TRIGGER}, \texttt{fetch*} is introduced for the policy to query the last reported value.\footnote{Different from {\tt attribute} which represents the current value of a device attribute, {\tt attribute}$^{*}$ denotes the latest reported value of this attribute.} The policy takes the action defined in \texttt{run} if the constraint defined by \texttt{branch} is satisfied; otherwise, the action in \texttt{else} is performed. Fields marked with ``[ ]'' are optional in the policy format.

\texttt{fetch*} and \texttt{else} are also introduced in \texttt{CHECK} items to handle similar situations. \mr{If the last reported temperature value to the platform is higher than 86, the policy does not need to report the newly queried temperature value to the platform (using \texttt{block()}) since $\mathbf{R}_{1}$'s condition-checking yields the correct result based on the last reported value. If the last reported temperature is equal to or lower than 86, the policy must report a temperature event higher than 86$^{\circ }F$ before reporting a ``present'' event to make sure the rule $\mathbf{R}_{1}$ is correctly executed.} Instead of reporting the real temperature value, the policy uses \texttt{randomize(86, MAX)} to report a random value. In $\mathbf{R}_{1}$, the temperature compares with a threshold (86$^{\circ }F$), so a random value between 86$^{\circ }F$ and the upper limit of a temperature \texttt{MAX} is sufficient for the condition checking.
\texttt{MAX}/\texttt{MIN} denotes the upper and lower boundary values of a specific attribute.\footnote{For instance, temperature (-460$\sim$10000 $^{\circ}\rm F$), humidity (0$\sim$100\%), luminance (0$\sim$100000 $\rm lux$).} We obtain such information from SmartThings Capabilities Reference \cite{smartthings2020capability}. 

With the above design, the policy derived from $\mathbf{R}_{1}$ (see Figure~\ref{fig_policy_from_rule}) expresses multi-faceted information for \tool to process data:

\begin{enumerate}[leftmargin=*]
    \item Context: when and only when an incoming event of $ps_{1}$ is ``present'', the current reading of $ts_{1}$ is higher than 86$^{\circ }F$, and the state of $f_{1}$ is not ``ON,'' \emph{some} data will be reported. Otherwise, the policy will be skipped and no data will be reported at all.
    \item \texttt{TRIGGER} data reporting: if the latest reported value of $ps_{1}$ is ``present,'' use the \texttt{diffKeep()} method to process the current value for reporting; otherwise, use \texttt{keep()}. 
    \item \texttt{CHECK} data reporting: if the latest reported value of $ts_{1}$ is higher than 86$^{\circ }F$, use the \texttt{block()} method to process the current value of $ts_{1}$; otherwise, use \texttt{randomize(86, MAX)}. 
\end{enumerate}


We implement a policy engine to parse and execute the policies. The policy engine listens to all the incoming raw data from the IoT devices and time-related information if registered. When receiving a new data datum $\texttt D$, the engine uses $\texttt D$ to evaluate the maintained policies one by one. Algorithm \ref{alg_policy_engine} summarizes the general workflow of how the engine evaluates and executes a policy $\mathbf{P}$. Specifically, it first checks if $\tt D$ matches the \texttt{type}, \texttt{subject}, and \texttt{attribute} in \texttt{TRIGGER}, and then examines if the value of $\texttt D$ satisfies the constraint specified by \texttt{operator} and \texttt{value}. If true, $\mathbf P$ is triggered and proceeds to execution. Then the engine evaluates all items specified in \texttt{CHECK}. Since the data required for evaluating the \texttt{CHECK} items are not new events but the current smart home status (e.g., the device working status), the policy engine fetches the information indexed by \texttt{type}, \texttt{subject} and \texttt{attribute} from a database \texttt{DB}, which stores the current states of all connected devices and updates them when devices report any change. Only when constraints defined in all \texttt{CHECK} items are satisfied, the actions defined in all \texttt{run} or \texttt{else} fields will be performed. During the above process, a policy terminates if there is any event mismatch or constraint violation. Besides, the policy engine also maintains another database \texttt{DB}$^{*}$ to keep record of the last reported data for each device attribute. 

\begin{algorithm}
	\scriptsize
	\SetKwInOut{Inputs}{Input}\SetKwInOut{Output}{Output}
	\Inputs{$\tt D \leftarrow$ new datum, $\tt P \leftarrow$ A privacy policy \\ $\tt DB\leftarrow$ Newest Device Status Database \\ $\tt DB^{*}\leftarrow$ Newest Reported Data Database}
	\Output{Reported Data Set $\tt DS$}
	
	\If{match($\tt D.source$, $\tt P.TRIGGER.(type, subject, attribute)$) $\tt and$ satisfy($\tt D.value$, $\tt P.TRIGGER.(operator, value)$)} {
		\ForEach{ $\tt checkitem\in\tt P.CHECK$ } {
			{$\tt val$} $\leftarrow$ {\it fetch}($\tt DB$, ${\tt checkitem}.(\tt type, subject, attribute)$) \\
			\If{!{\it satisfy}($\tt val$, $\tt checkitem.(\tt operator, value)$)} {
				{$\tt return$}\\
			}
			\If{!$\tt checkitem$.hasField({\tt [run]})} {
				{$\tt continue$}\\
			}			
			\eIf{ $\tt checkitem$.hasField({\tt [branch]}) } {
    		    $\tt val^{*} \leftarrow${\it fetch}($\tt DB^{*}$, ${\tt checkitem}.(\tt type, subject, attribute)$) \\
    	    	\eIf{satisfy($\tt val^{*}$, $\tt checkitem.(operator,value)$)} {
    			    $\tt DS \leftarrow$ {\bf run} $\tt checkitem.(method, paras)$\\
    		    } 
                {
    		        $\tt DS \leftarrow$ {\bf run} $\tt checkitem.(method1, paras1)$\\
    		    }
		    }
            {
    		    $\tt DS \leftarrow$ {\bf run} $\tt checkitem.(method, paras)$\\
    		}
		}
		\eIf{ $\tt P.TRIGGER$.hasField({\tt [branch]}) } {
		    $\tt val^{*} \leftarrow${\it fetch}($\tt DB^{*}$, ${\tt P.TRIGGER}.(\tt type, subject, attribute)$) \\
	    	\eIf{satisfy($\tt val^{*}$, $\tt P.TRIGGER.(operator1,value)$)} {
			    $\tt DS \leftarrow$ {\bf run} $\tt P.TRIGGER.(method, paras)$\\
		    } 
            {
		        $\tt DS \leftarrow$ {\bf run} $\tt P.TRIGGER.(method1, paras1)$\\
		    }
	    }
        {
		    $\tt DS \leftarrow$ {\bf run} $\tt P.TRIGGER.(method, paras)$\\
		}		
		
	} 
	\caption{The algorithm for executing a policy}
	\label{alg_policy_engine}
\end{algorithm}

\begin{figure}[t]
	\centering
	\includegraphics[width=0.35\textwidth]{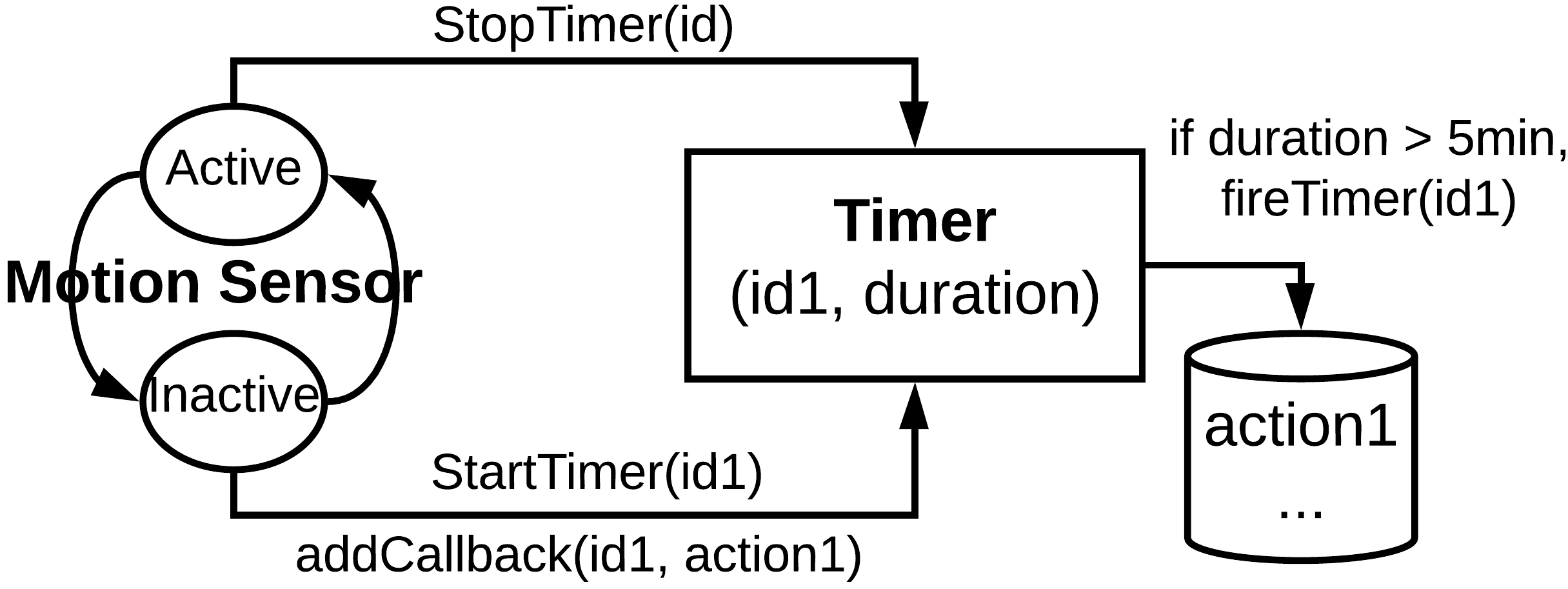}\\
	\caption{An example showing how \tool handles a rule whose condition has a timer. The methods are shown in Table~\ref{table_methods}. \texttt{action1} is defined to report ``inactive'' to the platform with method \texttt{keep} or \texttt{diffKeep}. Each timer maintains a list of actions which will be called when the timer fires. To accommodate, the timer in the rule is removed to avoid doubling the timer.}\label{fig_timer_example}
\end{figure}

\tool also deals with time-related automation. For instance, if a rule is defined as ``when the door is opened if time is after 18:00, turn on TV'', the derived policy needs to fetch system time for condition checking. When a rule has a timer, e.g., ``when motion sensor becomes inactive for 5 minutes, turn off the light'', multiple policies are bundled to operate by calling the methods for starting, stopping and firing a timer. Fig.~\ref{fig_timer_example} illustrates the workflow of how \tool handles this example.

\mr{The policy design supports all the rule features (e.g., state-checking, timers, delayed actions) employed by most popular platforms such as HomeKit, SmartThings, Alexa, Google Home, IFTTT, etc. However, the policy design currently does not support some uncommon automation rule semantics that are rarely seen. For instance, if a rule's trigger or condition checks historical values of a device (e.g., 
``\emph{if the precipitation in the past seven days is less than 100ml (\emph{trigger}), turn on the sprinkler system (\emph{action})}''),
\tool will simply let all the values pass to the platform to ensure the correctness of rule executions. The policy design of \tool is extendable to support such rules when needed. }

\subsubsection{User-Specified Policies}
\label{section_up}
We propose an interactive approach for users to specify data-protection policies. This is motivated by three reasons: 1) users have individual privacy preferences that cannot be derived from automation rules; for example, users might prioritize privacy rather than automation functionality for some device types during a time period or under certain situations; 2) the platform may integrate a third-party service but there is no rule extractor available to extract semantics from it; 3) users have rights to control the use of their data. In principle, UPs have higher priority than APs in controlling data. 

\begin{figure}[t]
	\centering
 \subfigure[]{
    \label{fig_info} 
    \includegraphics[width=0.22\textwidth]{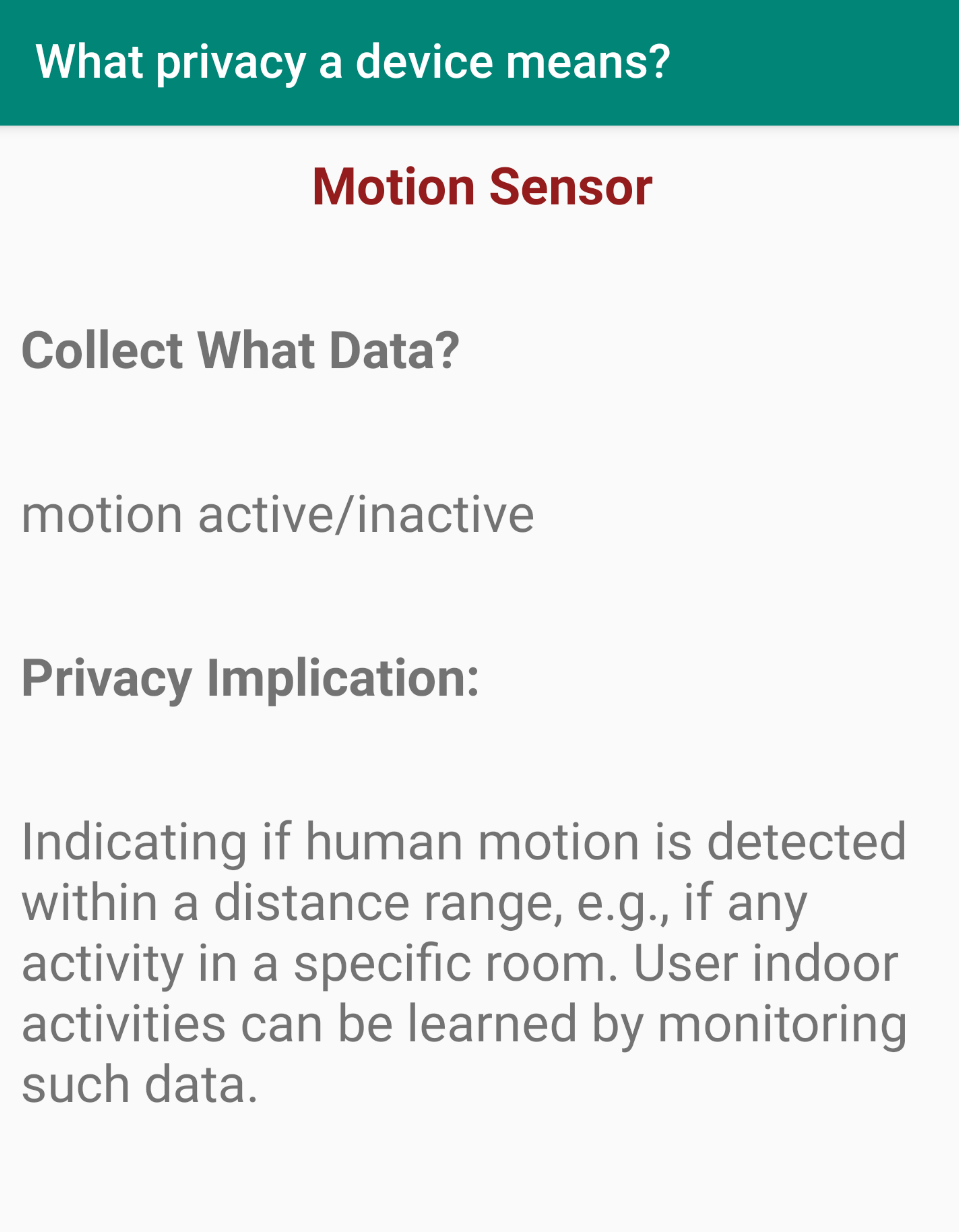}}
    \quad
 \subfigure[]{
    \label{fig_policy}
    \includegraphics[width=0.22\textwidth]{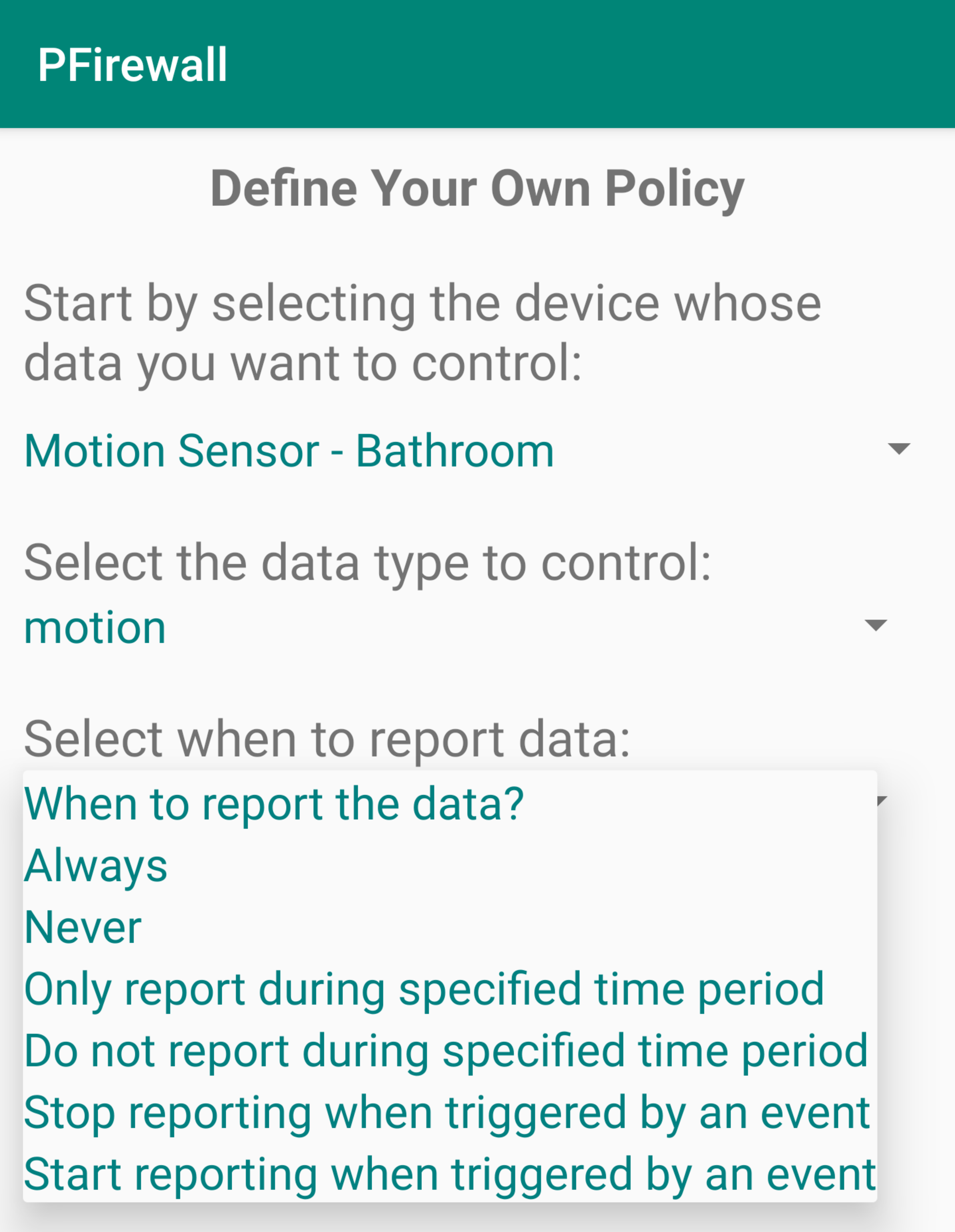}}
  \caption{Screenshots of \tool mobile app. The app provides an information tab showing users what data every device type generates and the corresponding privacy implications, and a policy tab allows users to define context-aware data control policies.}
  \label{fig_screenshot_app}
	\vspace{-10pt}
\end{figure}

We develop a mobile app for end-users to specify policies. As shown in Fig.~\ref{fig_info}, information is displayed to help users understand what privacy issues each device and its data may imply. With the templates in Fig.~\ref{fig_policy}, users are able to configure whitelist, blacklist and conditional-style policies during a specified time period or under certain contexts. Finally, UPs are encoded into the policy format in Listing~\ref{listing_policy} for execution.

\subsubsection{Policy Conflicts}
\label{section_policy_conflict}
It is possible that a user's UPs conflict with existing APs and hinder the automation since UPs are designed to override APs. Nevertheless, users need a warning that shows them what conflicts are caused and which automation rules are affected. Therefore, an automated policy conflict detection is needed. Two policies $\mathbf{P}_{1}$ and $\mathbf{P}_{2}$ \emph{conflict} with each other if all the following requirements are satisfied: 
(1) $\mathbf{P}_{1}$ and $\mathbf{P}_{2}$ are triggered simultaneously, i.e., an event makes both constraints $c_{T}^{1}$ and $c_{T}^{2}$ (defined in \texttt{TRIGGER} fields of $\mathbf{P}_{1}$ and $\mathbf{P}_{2}$, respectively) hold;
(2) both policies are finally executed, i.e., all the constraints $c_{i}^{1}$ and $c_{i}^{2}$ in the \texttt{CHECK} fields of both policies are evaluated true;
(3) two policies define different actions (i.e., data processing methods, parameters, or delays) for the same data. Formally, let $\mathbf{S}(C)$ denote the set of all possible contexts that satisfy the set of constraints $C$, and $\mathbf{O}(a)$, $\mathbf{E}(a)$ denote the object (i.e., the controlled data) and effects of a certain action $a$ (defined in both \texttt{TRIGGER} and \texttt{CHECK} fields). A conflict occurs when the formulas hold.

\vspace{-1.5em}
\begin{eqnarray}
\footnotesize
\left\{\begin{aligned}
&\mathbf{S}(c_{T}^{1}) \cap \mathbf{S}(c_{T}^{2}) \neq \emptyset,  \nonumber \\
&\mathbf{S}(c_{1}^{1}, c_{2}^{1}, \cdots) \cap \mathbf{S}(c_{1}^{2}, c_{2}^{2}, \cdots) \neq \emptyset, \\
&\exists i, j, \mathbf{O}(a_{i}^{1}) = \mathbf{O}(a_{j}^{2}), \mathbf{E}(a_{i}^{1}) \neq \mathbf{E}(a_{j}^{2}).
\end{aligned}
\right.
\label{equation_conflict}
\end{eqnarray}

\mr{\tool detects policy conflict for each newly submitted/updated UP against every AP. When an AP is added/updated due to automation rule adding/updating, \tool also detects it against all existing UPs.} To verify the first two formulas, we encode each constraint in a policy into a quantifier-free first-order formula: 
\begin{align*}
\footnotesize
\underbrace{(\texttt{type}[.\texttt{subject}[.\texttt{attribute}]])}_\text{data source and type} (\texttt{operator}) (\texttt{value}).
\end{align*}
To verify the last formulas, we check whether the two policies perform contradictory actions (by looking at the methods and parameters in \texttt{run} and \texttt{else} fields) on the same data.

Thus, the conflict detection is transformed into a constraint satisfaction problem that can be solved by a constraint programming solver. In our implementation, we use a JavaScript linear solver {\setttsize{\small}\texttt{javascript-lp-solver}} \cite{lpsolver}. \mr{If the constraint set derived from two policies is not solvable, it means the two policies have no conflict. Otherwise, two policies have a conflict.} \tool presents detected conflicts to users for decision making.  

\mr{The solving complexity is determined by the constraints derived from policies, which only involve simple numerical relationships and the search space (e.g., the range of device values, the number of devices in a policy) is very small. In our evaluation, it takes less than 400 milliseconds to check a pair of policies and we did not encounter a failure due to timeout.}

\subsection{Data Flow Mediation}
\label{section_relay}
\begin{figure}[tb]
	\centering
	\includegraphics[width=0.45\textwidth]{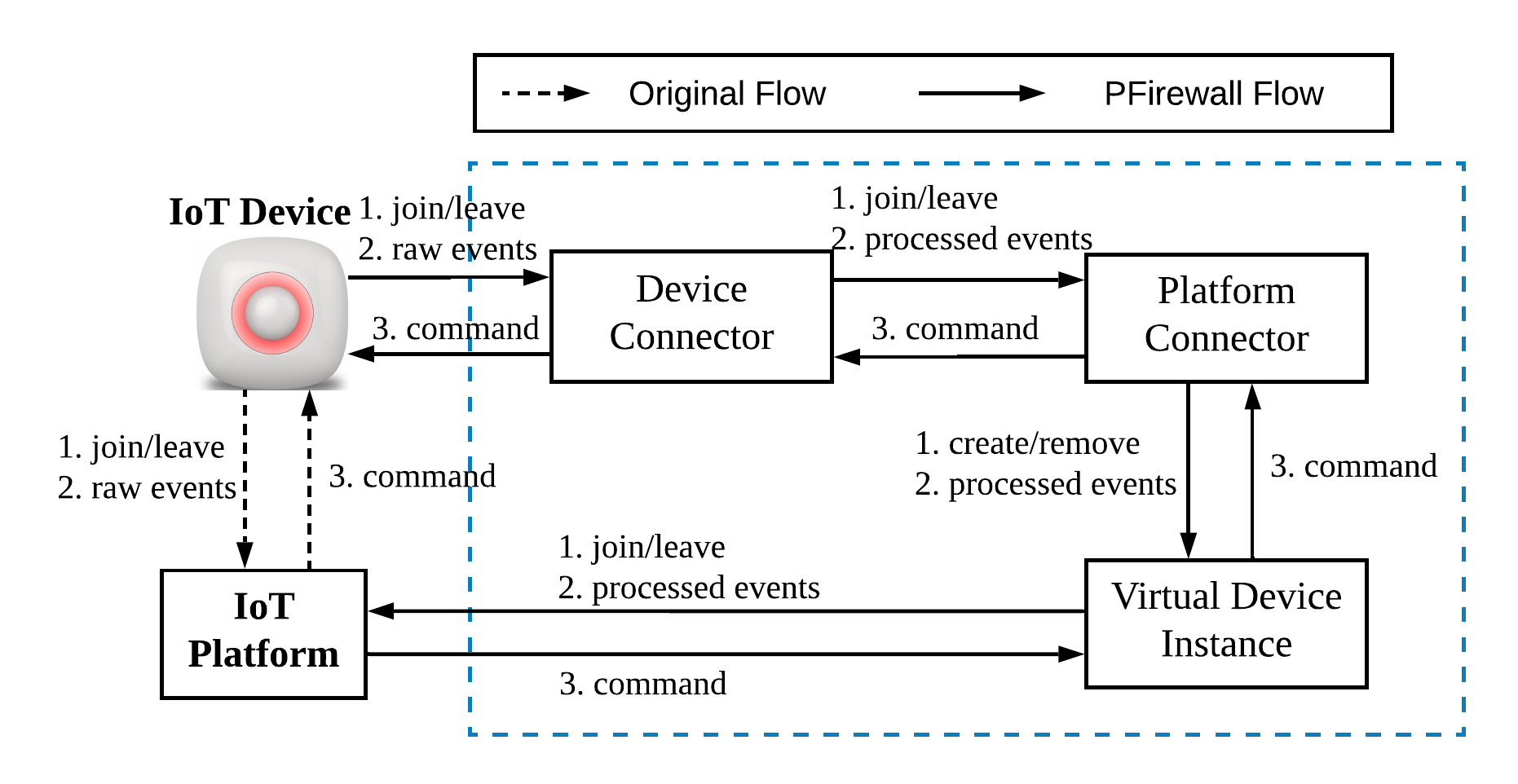}\\
	\caption{Workflow of the mediator.}\label{fig_gateway_virtual}
\end{figure}

To enforce data flow policies in a closed-source IoT system, we introduce a \emph{mediator} for relaying communications between IoT devices and the IoT platform, as shown in Fig. \ref{fig_gateway_virtual}. The mediator needs to act as both a device connector to interact with IoT devices and a platform connector, which generates a virtual device on behalf of each real device, to interact with the target platform. \mr{An advantage of our design is that the mediator \tool does not modify the IoT devices, hubs, or the platforms.}

\subsubsection{Connecting IoT Devices}
\label{section_device_connector}
\begin{figure}[tb]
	\centering
	\includegraphics[width=0.37\textwidth]{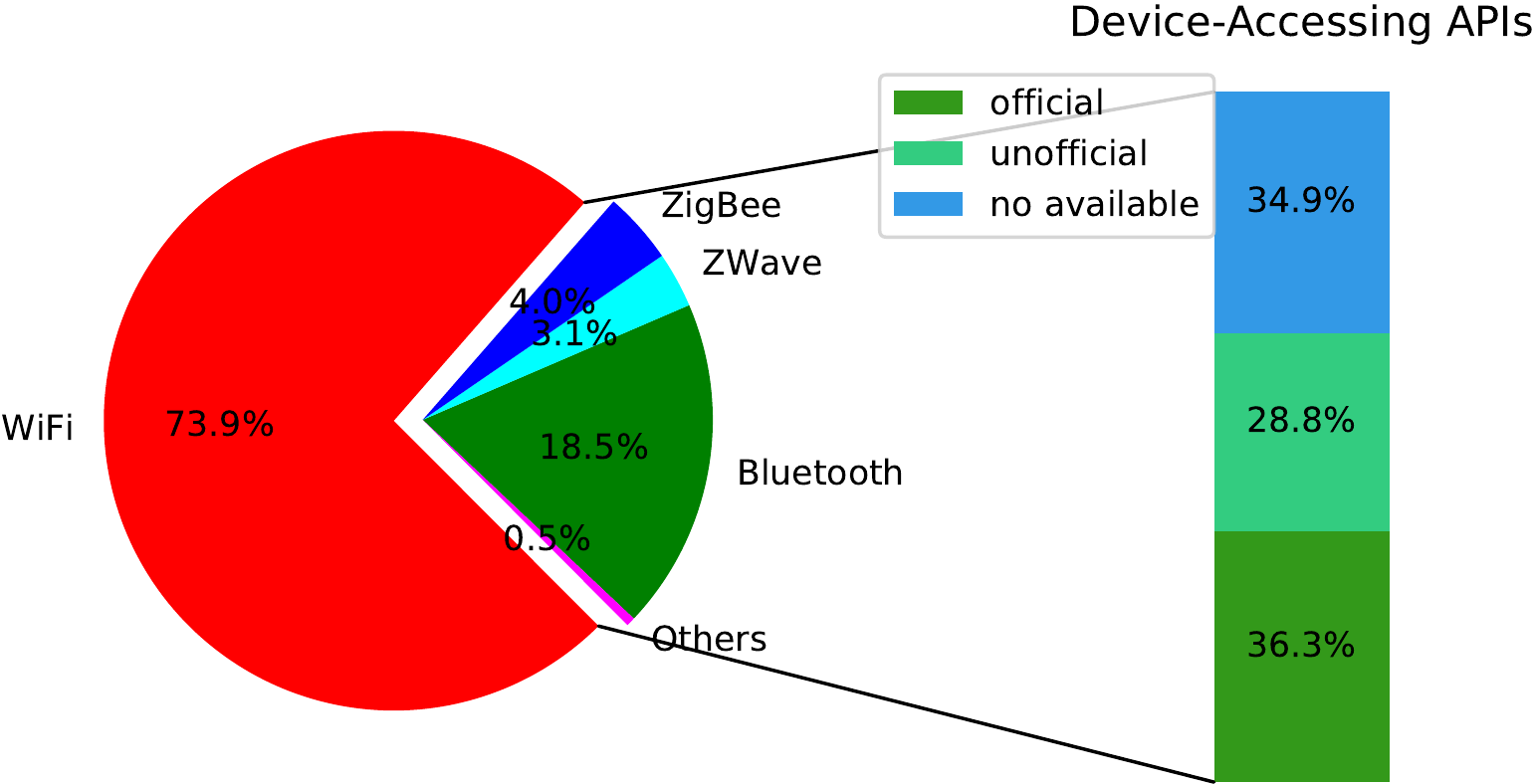}\\
	\caption{Statistics of popular communication technologies employed by commercial IoT devices and statistics of availability of device accessing APIs of WiFi devices. ``Others'' include devices that use other technologies including Clear Connect RF \cite{lutronclear}, HAP \cite{hap}, GSM \cite{gsm}. \tool could handle over 74\% of the investigated devices (including WiFi-connected devices with unofficial APIs).}\label{fig_device_investigation}
	\vspace{-10pt}
\end{figure}

\mr{We investigate the communication technologies used by IoT devices in the market, i.e., the top 1000 IoT devices with the most total reviews on five popular online stores (Amazon, BestBuy, Costco, Walmart, and Home Depot). Our investigation focuses on two important questions: (1) the portion of IoT devices that use each of the popular wireless technologies and (2) among the devices using end-to-end encryption technologies (such as WiFi) to connect with endpoint/vendor clouds, how many of them provide cloud APIs for accessing devices. The result is presented in Fig. \ref{fig_device_investigation}. Currently, \tool cannot handle IoT devices that use proprietary or end-to-end encrypted connections but do not provide accessing APIs.} However, we envision that more IoT cloud endpoints will provide open APIs in the future to increase interoperability and facilitate service integration. To demonstrate the feasibility, we choose two open-source standard protocols (ZigBee, Z-Wave) and the most common end-to-end encrypted protocols (WiFi connection with the vendor clouds which provide REST cloud APIs). 

To play the role of a device connector, the mediator needs to handle 3 major tasks associated with IoT devices: 1) adding or removing devices to \tool; 2) receiving events from devices; and 3) sending commands to devices. The device connector alike functions have been implemented and are available in many open-source platforms, e.g., openHAB \cite{openhab2019} and Mozilla IoT \cite{mozillaiot2019}, which allow developers to add add-ons for integrating various IoT devices using different communication techniques. At the time of research, openHAB supports 275 bindings that have been tested to work with hundreds of commercial IoT devices and Mozilla IoT also has tested more than 100 mainstream devices. In our implementation, we utilize the ZigBee and Z-Wave add-ons of Mozilla IoT to realize connecting with ZigBee and Z-Wave devices. Specifically, the mediator is built on a Raspberry Pi with a Digi XStick USB dongle (ZB mesh version) and an Aeotec Z-Stick (Gen5) to extend ZigBee and Z-Wave capabilities, respectively. For WiFi devices, we use the REST APIs provided by three vendor clouds (Philips Hue, LIFX, and Honeywell) to access devices connected to them.

\subsubsection{Connecting Various Platforms} \label{section:connect_platform}
\mr{When a physical device is connected to \tool, the platform connector creates an instance of \emph{virtual device} for it. The virtual device interacts with a platform on behalf of the physical device, i.e., it
(1) is discovered by the platform as a new \emph{device instance}, (2) reports the post-filtering events to the platform, and (3) forwards commands from the platform to the physical device (via the device connector).}
Popular platforms support various connectivity protocols; for example, SmartThings classic supports LAN- and cloud-based device integration \cite{lanconnected2018}, the new SmartThings supports cloud-connected devices \cite{stDevices}, and openHAB supports the Message Queuing Telemetry Transport (MQTT) protocol \cite{mqtt-openhab2019}, Mozilla IoT provides REST-based Web Things framework and APIs \cite{webofthings2019}, and Wink allows creating REST API devices \cite{winkrestful2019}. 
These features allow a virtual device instance created by the platform connector to talk with the platform as if it is a real device.
We implement the mediator to work with three representative systems with different interfacing techniques: SmartThings classic, new SmartThings, and openHAB, \mr{without modifying the software of these platforms or the hubs manufactured by them.} 

\begin{figure}[tb]
	\centering
	\includegraphics[width=0.43\textwidth]{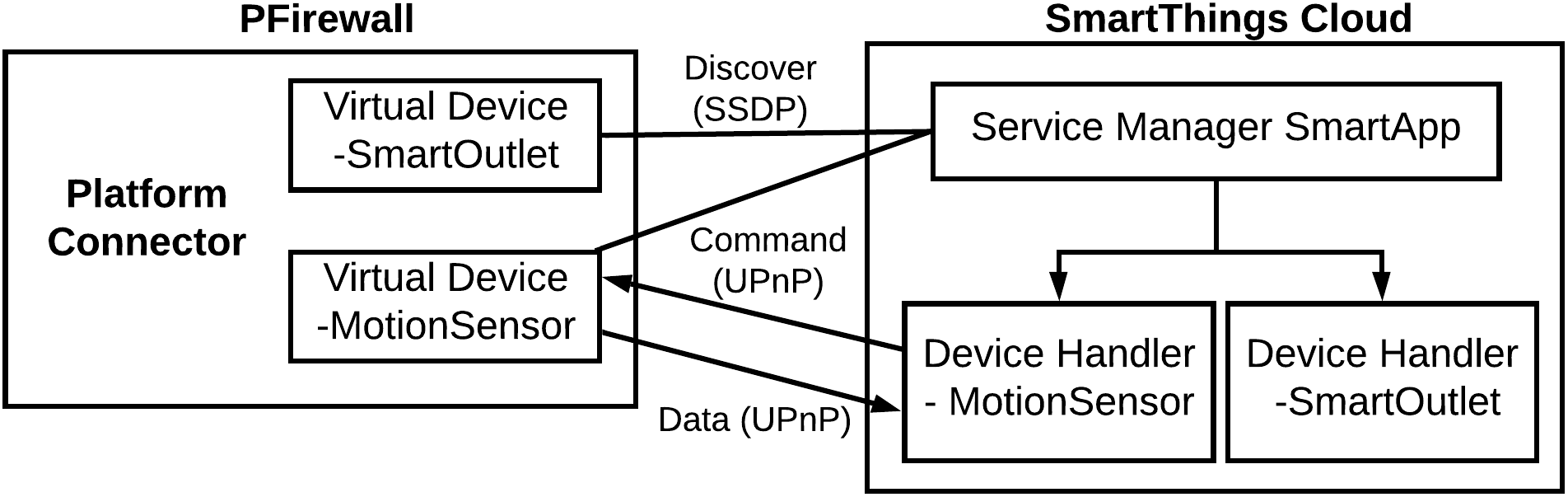}\\
	\caption{Overview of interfacing with SmartThings classic. Note that the communication between \tool and the SmartThings cloud is bridged by a SmartThings hub, which is omitted here.}\label{fig_upnp_implementation}
	\vspace{-10pt}
\end{figure}

{\setttsize{\small}
\vspace{4pt}
\noindent\textbf{Interfacing with SmartThings Classic System}\vspace{2pt} \\
We choose LAN as the protocol for communicating with the SmartThings classic system.
SmartThings classic uses a \texttt{device} \texttt{handler} (DH) for abstracting each supported device type. Accordingly, we build a virtual device (VD) type for each DH, as illustrated in Fig.~\ref{fig_upnp_implementation}. We develop a \texttt{service} \texttt{manager} SmartApp on SmartThings, which uses SSDP (Simple Service Discovery Protocol) to discover VD instances on the same LAN as the SmartThings hub. SmartThings classic uses a combination of IPs and ports to uniquely identify devices. To be discovered as different devices, each VD instance uses a unique port. After discovering a VD instance, the service manager app adds it as a \texttt{child} \texttt{device}. When a \texttt{child} \texttt{device} is added, SmartThings classic automatically selects a DH and abstracts it according to the \texttt{model} property of the \texttt{child device}. Hence, we make the \texttt{model} of a VD instance the same as the \texttt{name} of the target DH. After the initial connection, a VD instance on the mediator side interacts with a DH instance on the SmartThings via the UPnP (Universal Plug and Play) protocol. In additional, we adapt all DHs for ZigBee/Z-Wave devices available in the SmartThings IDE. For each DH, we add a \texttt{subscribe()} function that perform the SUBSCRIBE operation in UPnP. When a DH is instantiated (which means a VD instance is created and a child device is added), it uses the IP and port to send a SUBSCRIBE SOAP (Simple Object Access Protocol) message to the VD instance. Moreover, we replace the original code with new code for receiving and sending SOAP messages. The original code is in \texttt{parse} and command functions of DH for receiving ZigBee/Z-Wave data and sending ZigBee/Z-Wave commands. At this point, the VD and DH instances become addressable to each other and a subscribe/publish based UPnP communication is implemented to report data and send commands.
}

{\setttsize{\small}
\vspace{4pt}
\noindent\textbf{Interfacing with the New SmartThings System}\vspace{2pt} \\
We achieve interfacing with the new SmartThings system by developing a SmartThings schema connector \cite{stSchemaConnector} for \tool platform connector, which requires a cloud-to-cloud connection between \tool and SmartThings. To this end, we use \emph{fast reverse proxy} \cite{frp} to expose \tool (placed in a home network) to the SmartThings cloud, register \tool as an application in SmartThings Developer Workspace \cite{stworkspace}, and make the platform connector (for the new SmartThings) run as an HTTPs server. The \tool schema connector implements handlers for the interaction types \cite{interactiontypes} pre-defined by SmartThings, including \texttt{Reciprocal} \texttt{Access} \texttt{Token} (mutual authentication), \texttt{Discovery} (SmartThings discovers devices connected to \tool), \texttt{State} \texttt{Refresh} (SmartThings queries device states from \tool), \texttt{Command} (SmartThings sends commands to \tool), \texttt{Device} \texttt{State} \texttt{Callback} (\tool reports events to SmartThings), \texttt{Discovery Callback} (\tool reports newly added devices to SmartThings). For privacy reasons, when the new SmartThings system sends a \texttt{State Refresh}) request, \tool reports the previously-reported states of the requested devices (instead of reporting the current device states).
}

\begin{figure}[tb]
	\centering
	\includegraphics[width=0.49\textwidth]{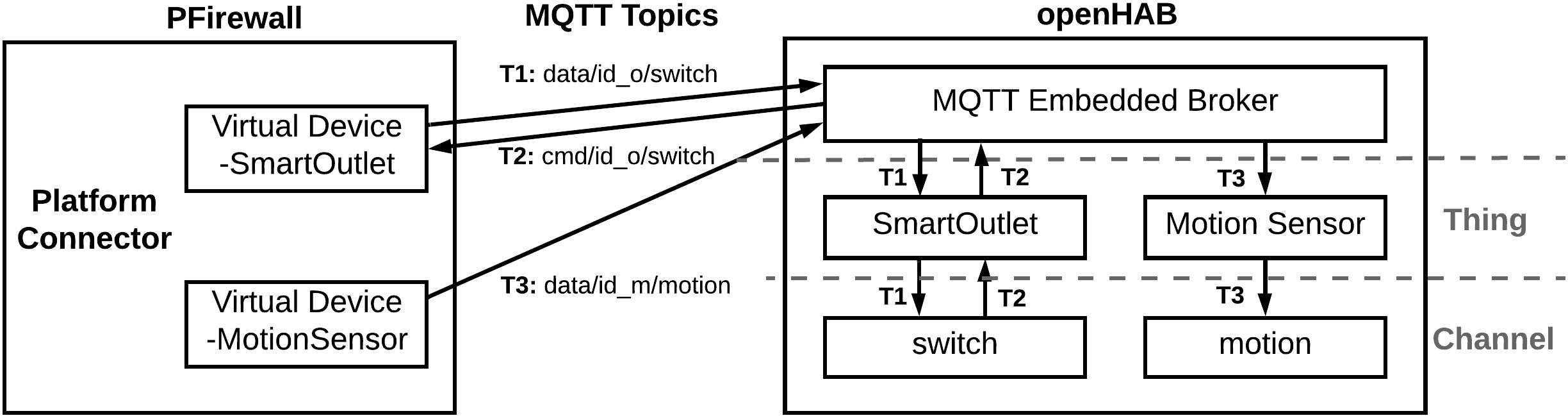}\\
	\caption{Overview of interfacing with openHAB.}\label{fig_mqtt_implementation}
	\vspace{-10pt}
\end{figure}

\vspace{4pt}
\noindent\textbf{Interfacing with openHAB}\vspace{2pt} \\
\label{appendix_openhab_implementation}
\tool uses MQTT to interface with openHAB because it is a general connectivity protocol and allows for virtualizing any device types with flexibility. Fig.~\ref{fig_mqtt_implementation} shows the high-level architecture of the integration. openHAB provides an embedded MQTT broker.
When a new device is added to \tool, a VD instance is created. For each supported attribute of this devices, if it is a read-only attribute (e.g., motion, temperature), the VD subscribes to a topic {\setttsize{\small}\texttt{data/{\{}deviceId{\}}/{\{}attribute{\}}}}
for publishing events; if it is a writable attribute (e.g., switch), the VD subscribes to two topics {\setttsize{\small}\texttt{data/{\{}deviceId{\}}/{\{}attribute{\}}}} and {\setttsize{\small}\texttt{cmd/{\{}deviceId{\}}/{\{}attribute{\}}}} for publishing events and receiving commands, respectively.
Meanwhile, by creating a new entry in openHAB thing configuration file \cite{ohthingconfiguration}, \tool adds a corresponding MQTT thing instance with channels to openHAB. Each thing channel in openHAB corresponds to an attribute in VD (e.g., motion, switch) and shares the same topic(s). Thus, the corresponding VD in \tool and thing channels in openHAB could exchange events and commands via the shared MQTT topic(s).

\section{Evaluation} 
\subsection{Evaluation Setup}
\begin{figure}[tb]
  \centering
  \subfigure[T$_{1}$]{
    \label{fig_t1}
    \includegraphics[width=0.23\textwidth]{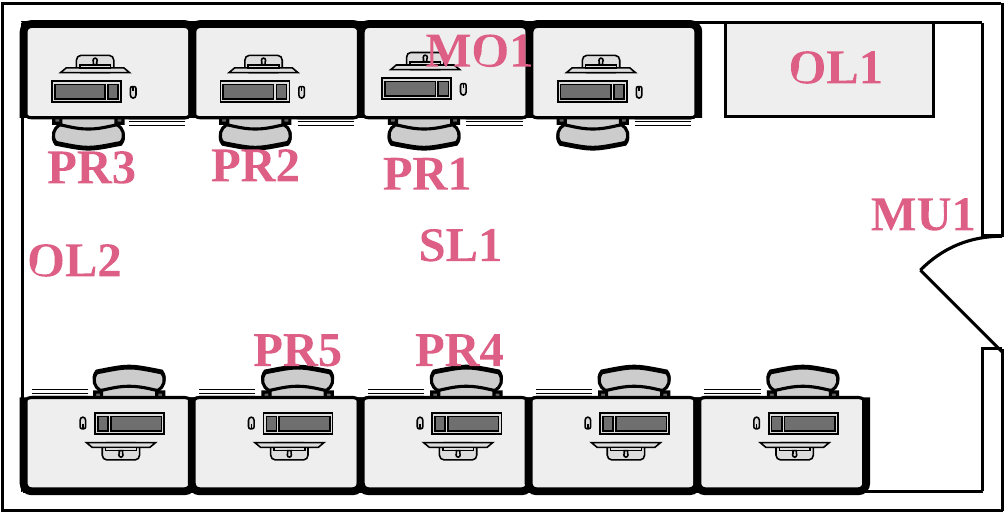}}
  \subfigure[T$_{2}$]{
    \label{fig_t2}
    \includegraphics[width=0.225\textwidth]{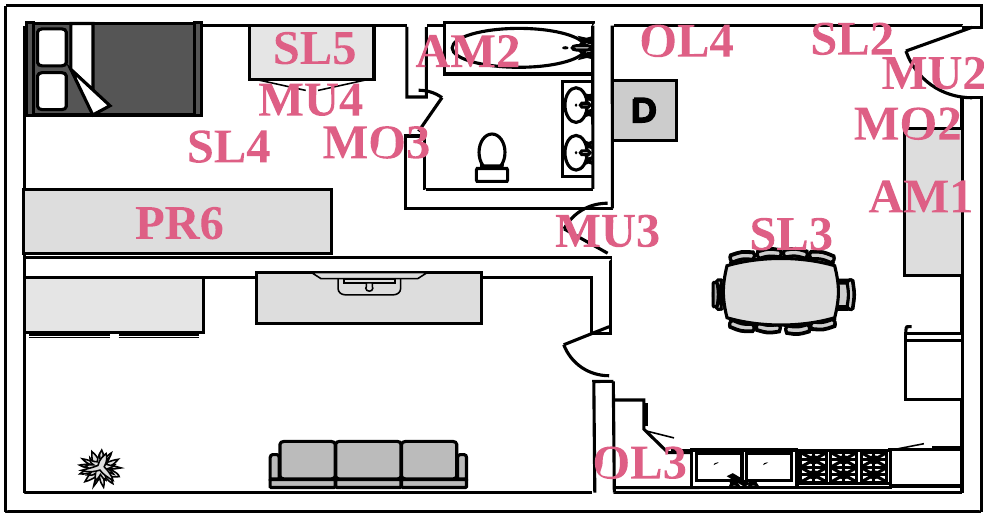}} \\
  \subfigure[T$_{3}$]{
    \label{fig_t3}
    \includegraphics[width=0.22\textwidth]{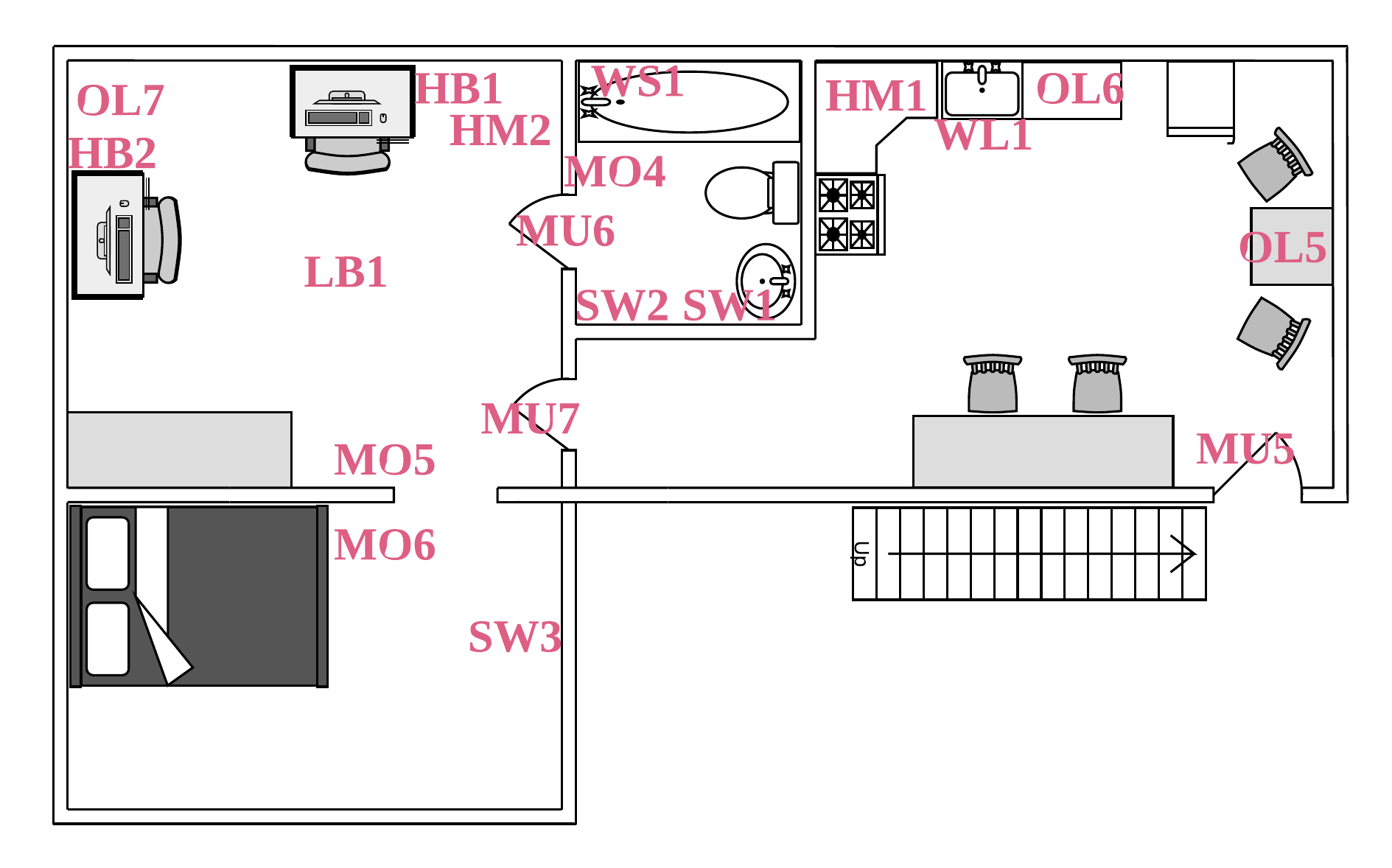}}
  \subfigure[T$_{4}$]{
    \label{fig_t4}
    \includegraphics[width=0.23\textwidth]{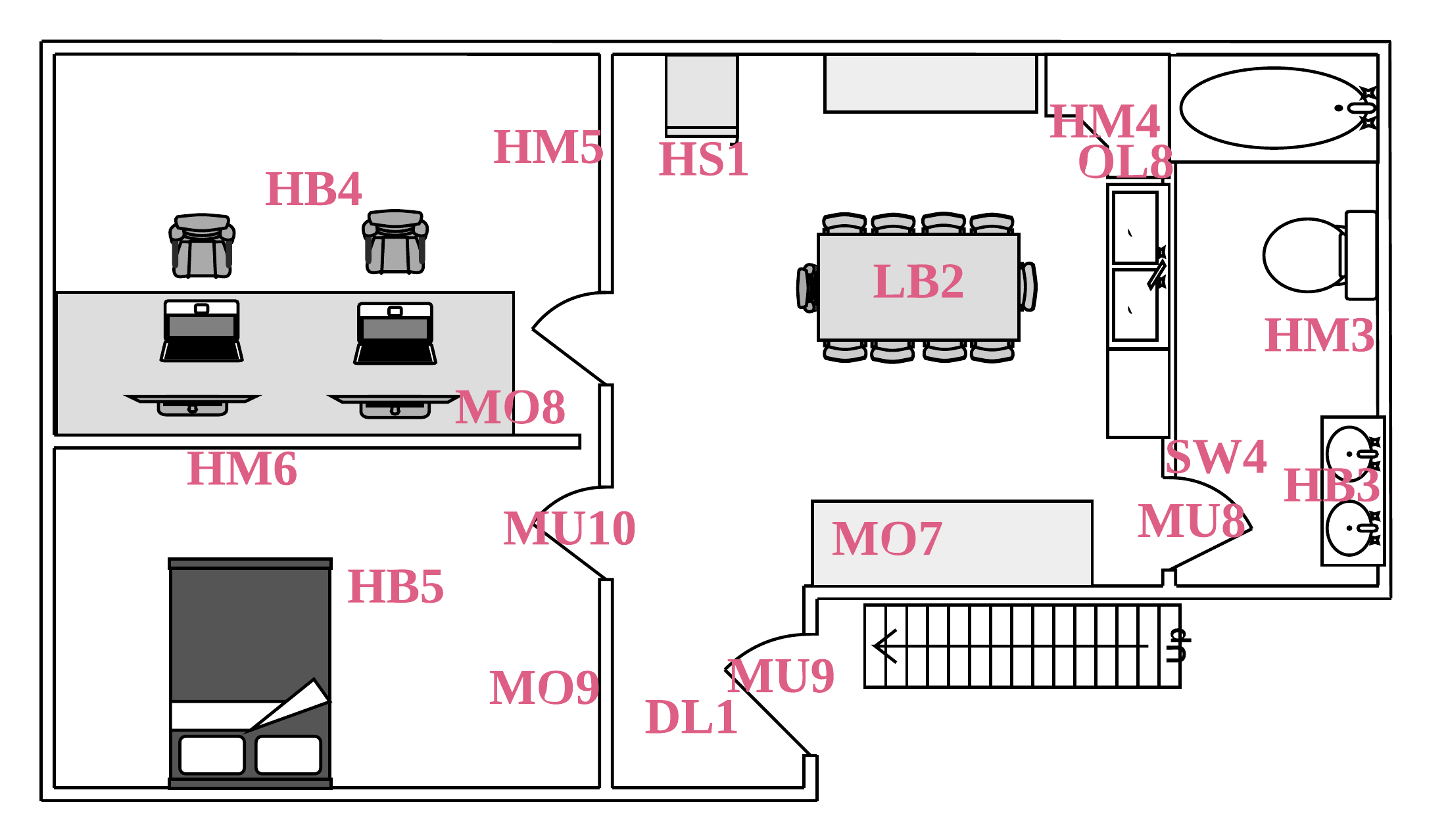}}
  \caption{The layout and device placement in four testbeds. Each presence sensor \texttt{PR1}$\sim$\texttt{6} is carried with a member. }
  \label{fig_setup}
  \vspace{-10pt}
\end{figure}

We set up four real-world testbeds for evaluating the performance of \tool: an office with 5 members (T$_{1}$), a two-bedroom apartment with 1 member (T$_{2}$), and two one-bedroom apartments each of which has 2 members (T$_{3}$ and T$_{4}$).\footnote{We have received the IRB approval (See Appendix~\ref{irb}).}
The floor plan and device placement are shown in Fig.~\ref{fig_setup}. The deployed platform and device types are given in Table~\ref{table_devices}. The configured automation rules of the testbeds are listed in Table~\ref{table_rules}. We ask the members of the testbeds to write down their desired automation rules in English for their own office or home, and then the authors implement the specified rules by installing official apps (T$_{1}$, T$_{2}$), developing custom apps (T$_{1}$, T$_{2}$, T$_{4}$), and/or using the mobile companion app interface (T$_{3}$). The rules are tested and adjusted for one week to make sure they meet the needs of the testbed members.

\begin{table*}[tb]
	\centering
	\scriptsize
	\caption{Deployed \mr{platforms} and devices in the four testbeds. \tool uses ZigBee (\texttt{MU}, \texttt{MO}, \texttt{OL}, \texttt{SL}, \texttt{WS}, \texttt{SW}), Z-Wave (\texttt{AM}), LAN (\texttt{PR}) or cloud APIs (\texttt{WL}, \texttt{HM}, \texttt{HB}, \texttt{LB}, \texttt{HS}) to connect these devices. }
	\renewcommand\arraystretch{1}
	\newcommand{\tabincell}[2]{\begin{tabular}{@{}#1@{}}#2\end{tabular}}
	\newcolumntype{P}[1]{>{\raggedright\arraybackslash}p{#1}}
	\newcolumntype{M}[1]{>{\centering\arraybackslash}m{#1}}
	\begin{tabular}{|M{0.9cm}|M{1.3cm}|P{3.1cm}|c|c|M{0.4cm}|}	
	\hline
		\bf Testbed & \bf \mr{Platform} & \bf Device ({\tt Abbr.}) & \bf \mr{Protocol}  & \centering\bf Attribute & \bf Num\\\hline	 
		\multirow{6}*{T$_{1}$} & \multirow{6}*{\tabincell{c}{SmartThings\\Classic}}
		& SmartThings hub v2 & ZigBee, Z-Wave, WiFi, Ethernet &  -- & 1 \\ 
		~ && Multipurpose sensor ({\tt MU}) & ZigBee & contact, temp. & 1 \\
		~ && Motion sensor ({\tt MO}) & ZigBee & motion, temp. & 1 \\
		~ && Smart outlet ({\tt OL}) & ZigBee & switch, power & 2 \\
		~ && Smart light ({\tt SL}) & ZigBee & switch & 1  \\
		~ && Presence sensor ({\tt PR}) & ZigBee & presence & 5\\
		\hline
		\multirow{7}*{T$_{2}$} & \multirow{7}*{\tabincell{c}{SmartThings\\Classic}}
		&	SmartThings hub v2 &  ZigBee, Z-Wave, WiFi, Ethernet & --  & 1 \\
		~ && Multipurpose sensor ({\tt MU}) & ZigBee & contact, temp. & 3\\
		~ && Motion sensor ({\tt MO}) & ZigBee & motion, temp. & 2\\
		~ && Smart outlet ({\tt OL}) & ZigBee & switch, power & 2 \\
		~ && Smart bulb ({\tt SL}) & ZigBee & switch & 4  \\
		~ && Aeotec MultiSensor ({\tt AM}) & Z-Wave & motion, humidity, luminance & 2  \\
		~ && Presence sensor ({\tt PR}) & ZigBee & presence & 1\\
		\hline
		\multirow{9}*{T$_{3}$} & \multirow{9}*{\tabincell{c}{SmartThings\\New System}}
		&	Philips Hue Bridge &  ZigBee, Ethernet & --  & 1 \\
		~ && Multipurpose sensor ({\tt MU}) & ZigBee & contact, temp. & 3\\
		~ && Motion sensor ({\tt MO}) & ZigBee & motion, temp. & 3\\
		~ && Smart outlet ({\tt OL}) & ZigBee & switch, power & 3 \\
		~ && Smart switch ({\tt SW}) & ZigBee & switch & 3  \\
		~ && Water sensor ({\tt WS}) & ZigBee & water, temp. & 1  \\
		~ && Honeywell water leakage ({\tt WL}) & WiFi & water, temp. & 1  \\
		~ && Hue motion sensor ({\tt HM}) & ZigBee & motion, temp., luminance & 2\\
		~ && Hue bulb ({\tt HB}) & ZigBee & switch & 2\\
		~ && LIFX bulb ({\tt LB}) & WiFi & switch & 1\\
		\hline
		\multirow{9}*{T$_{4}$} & \multirow{9}*{openHAB}
		& Philips Hue Bridge &  ZigBee, Ethernet & --  & 1 \\
		~ && Multipurpose sensor ({\tt MU}) & ZigBee & contact, temp. & 3\\
		~ && Motion sensor ({\tt MO}) & ZigBee & motion, temp. & 3\\
		~ && Smart outlet ({\tt OL}) & ZigBee & switch, power & 1 \\
		~ && Smart switch ({\tt SW}) & ZigBee & switch & 1  \\
		~ && Hue motion sensor ({\tt HM}) & ZigBee & motion, temp., luminance & 4\\
		~ && Hue bulb ({\tt HB}) & ZigBee & switch & 3\\
		~ && Hue switch ({\tt HS}) & ZigBee & button & 1\\
		~ && LIFX bulb ({\tt LB}) & WiFi & switch & 1\\
		~ && Dimmable light ({\tt DL3}) & ZigBee & level & 1\\
	\hline
	\end{tabular}
	\label{table_devices}
\end{table*}

\begin{table*}[tb]
	\centering
	\scriptsize
	\caption{Installed rules in all testbeds. Source: official app (O), custom app (C), templates on mobile/web interface (T).} 
	\renewcommand\arraystretch{1}
	\newcommand{\tabincell}[2]{\begin{tabular}{@{}#1@{}}#2\end{tabular}}
	\newcolumntype{P}[1]{>{\raggedright\arraybackslash}p{#1}}
	\newcolumntype{M}[1]{>{\centering\arraybackslash}m{#1}}
	\begin{tabular}{|M{0.65cm} |M{0.3cm}|M{0.6cm}| P{14.5cm}|}	
	\hline
		\bf Testbed & \bf ID & \bf Source  & \bf Description and Device Bindings  \\\hline	 
		\multirow{6}*{\tabincell{c}{T$_{1}$}}
		~ &	1  & O & When front door ({\tt MU1}) is opened, turn on light ({\tt SL1}).  \\ 
		~ & 2 & O & When no motion ({\tt MO1}) or presence ({\tt PR1$\sim$5}) is detected for 5 minutes, turn off light ({\tt SL1}).    \\
		~ & 3 & C & When presence ({\tt PR1}) becomes present, if time is before 12am, turn on coffee machine ({\tt OL1}). \\
		~ & 4 & C & When motion ({\tt MO1}) detected, if temperature ({\tt MU1}) is below 70$^{\circ} F$, turn on heater ({\tt OL2}). \\
		~ & 5 & C & When motion ({\tt MO1}) is active for longer than 60 minutes, send a message to alert.     \\
		~ & 6 & C & When door ({\tt MU1}) is open, if no presence ({\tt PR2$\sim$5}), send a message. \\
		\hline
		\multirow{8}*{\tabincell{c}{T$_{2}$}} 
		&	7 & O & When door is opened ({\tt MU2}), turn on light ({\tt SL2}).   \\
		~ & 8 & O & When motion ({\tt MO2}) active if luminance ({\tt AM1}) is below 15 LUX, turn on light ({\tt SL3}). \\
		~ & 9 & O & When presence ({\tt PR6}) becomes present if between 5-8pm, turn on oven ({\tt OL3}). \\
		~ & 10 & O & When motion ({\tt MO2}) active if not presence ({\tt PR6}), send a notification. \\
        ~ &	11 & O & When wardrobe door ({\tt MU4}) open, turn on light ({\tt SL5}); when door ({\tt MU4}) close, turn off light ({\tt SL5}). \\
		~ & 12 & O & When motion ({\tt MO2}), if temp. ({\tt MU3}) below 68$^{\circ} F$, turn on heater ({\tt OL4}); when motion ({\tt MO2}) inactive for 20 minutes, turn off heater ({\tt OL4}). \\
		~ & 13 & C & When door ({\tt MU3}) opened, turn on light ({\tt SL4}); when door ({\tt MU3}) closed if motion ({\tt MO3}) inactive for 5 minutes, turn off light ({\tt SL4}). \\
		~ & 14 & C & When humidity ({\tt AM2}) exceeds 85\% if motion ({\tt AM2}) active but motion ({\tt MO3}) keeps inactive for 30 minutes, send a notification. \\
		\hline
		\multirow{10}*{\tabincell{c}{T$_{3}$}}
		&	15 & T & When bedroom motion ({\tt MO6}) active between 10am and 12am, turn on light ({\tt SW3}).   \\
		~ & 16 & T & When kitchen motion ({\tt HM1}) active between 7am and 12am, turn on microwave outlet ({\tt OL6}). \\
		~ & 17 & T & When sink water leakage ({\tt WL1}) detected and all motion ({\tt MO4}$\sim${\tt6}, {\tt HM1}$\sim${\tt2}) inactive, send a text message. \\
		~ & 18 & T & When front door ({\tt MU5}) open and luminance ({\tt HM2}) is below 30 LUX, turn on lights ({\tt LB1}, {\tt OL7}). \\
        ~ &	19 & T & When bathroom motion ({\tt MO4}) active and water sensor ({\tt WS1}) wet, turn on fan ({\tt SW1}). \\
		~ & 20 & T & When 7am, turn on coffee outlet ({\tt OL5}); when 6pm, turn off coffee outlet ({\tt OL5}). \\
        ~ &	21 & T & When bathroom motion ({\tt MO4}) active between 10am and 1am, turn on light ({\tt SW2}). \\
		~ & 22 & T & When bathroom motion ({\tt MO4}) inactive and bathroom door ({\tt MU6}) closed, turn off light switch ({\tt SW2}) after 10 minutes. \\
		~ & 23 & T & When study room motion ({\tt MO5}) active between 9am and 12am and luminance ({\tt HM2}) is below 30 LUX, turn on lights ({\tt LB1}, {\tt OL7}, {\tt HB1}$\sim${\tt2}). \\
		~ & 24 & T & When study desk motion ({\tt HM2}) active between 9am and 12am and luminance ({\tt HM2}) is below 30 LUX, turn on light ({\tt LB1}, {\tt OL7}, {\tt HB1}$\sim${\tt2}). \\
		\hline
		\multirow{12}*{\tabincell{c}{T$_{4}$}}
		&	25 & C & When stovetop motion ({\tt HM4}) active, if Hue switch ({\tt HS1}) activates button 1 (day mode) or 3 (nap mode), turn on lamp outlet ({\tt OL8}). \\
		~ & 26 & C & When stovetop motion ({\tt HM4}) inactive for 5 minutes, turn off the lamp outlet ({\tt OL8}). \\
		~ & 27 & C & When living room motion ({\tt MO7}) active, if Hue switch ({\tt HS1}) activates button 1 (day mode) or 3 (nap mode), turn on living room light ({\tt LB2}). \\
        ~ &	28 & C & When living room motion ({\tt MO7}) inactive for 20 minutes, turn off living room light ({\tt LB2}). \\
		~ & 29 & C & When bathroom motion ({\tt HM3}) active, if Hue switch ({\tt HS1}) activates button 1 (day mode) or 3 (nap mode), turn on lights ({\tt SW4}, {\tt HB3}); if Hue switch ({\tt HS1}) activates button 2 (night mode), turn on light switch ({\tt SW4}). \\
		~ & 30 & C & When bathroom motion ({\tt HM3}) inactive and door ({\tt MU8}) open for 15 minutes, turn off lights ({\tt SW4}, {\tt HB3}). \\
		~ & 31 & C & When bedroom motion ({\tt HM6}) active, if Hue switch ({\tt HS1}) activates button 1 (day mode), turn on light ({\tt HB5}). \\
        ~ &	32 & C & When bedroom motion ({\tt HM6}) inactive for 10 minutes, turn off light ({\tt HB5}). \\
		~ & 33 & C & When study room motion ({\tt HM5}) active, if Hue switch ({\tt HS1}) activates button 1 (day mode) or 3 (nap mode), turn on light ({\tt HB4}). \\
		~ & 34 & C & When study room motion ({\tt HM5}) inactive for 30 minutes, turn off light ({\tt HB4}). \\
		~ & 35 & C & When front door ({\tt MU9}) open, set the dimmer ({\tt DL1}) brightness level to 100\%; after 5 minutes, set the dimmer to 0\%. \\
	\hline
	\end{tabular}
	\label{table_rules}
\end{table*}  

\subsection{Performance of Data Mediating}
\label{label_performance_mediating}
To test the correctness of \tool, we disable data filtering in each testbed, i.e., \tool simply forwards device events to the target platform without executing any policies. We use the built-in device-level event logs provided by SmartThings Groovy IDE \cite{smartthings2020ide} to capture the data received by SmartThings classic and the new SmartThings systems.
openHAB does not provide a similar web interface for looking up events, so we build an event logger by utilizing the openHAB REST API \cite{ohREST}. We observe duplicate events in the event logs of SmartThings, and we remove duplicate events before data analysis. 


We run experiments using the above setting in all four testbeds for one week and compare the event sequence received by \tool and the platform for each testbed. 
The experimental results are given in Table~\ref{table_mediating_result}, which demonstrate that our mediator works correctly in relaying IoT device events to the platforms.
 
\begin{table}[tb]
	\centering
	\scriptsize
	\caption{Comparison of events received by \tool mediator and the target platform in all testbeds. \textbf{Total}: the total data volume received by \tool mediator and the platform, respectively. Due to space limits, we list partial representative devices and attributes, covering different communication protocols and attribute types (binary/numeric-value) in every platform.
	}
	\renewcommand\arraystretch{1}
	\newcommand{\tabincell}[2]{\begin{tabular}{@{}#1@{}}#2\end{tabular}}
	\newcolumntype{P}[1]{>{\raggedright\arraybackslash}p{#1}}
	\newcolumntype{M}[1]{>{\centering\arraybackslash}m{#1}}
	\newcolumntype{R}[1]{>{\raggedleft\arraybackslash}p{#1}}
	\begin{tabular}{|M{1cm} |c| M{1.5cm}|M{1.5cm}|M{1.6cm}|}
	\hline
		\bf Testbed & \centering\bf Device  & \bf Attribute  & \bf Total & \bf Inconsistency \\\hline	 
		\multirow{6}*{\tabincell{c}{T$_{1}$}} 
		& {\tt MU1} & contact & 1372, 1372 & 0 \\
		~ & {\tt MU1} & temperature & 6174, 6174 & 0 \\
		~ & {\tt MO1} & motion & 1598, 1598 & 0\\
		~ & {\tt OL1} & switch & 24, 24 & 0\\
		~ & {\tt SL1} & switch & 18, 18 & 0 \\
		~ & {\tt PR1} & presence & 46, 46 & 0 \\
		\hline
		\multirow{3}*{\tabincell{c}{T$_{2}$}}
		 & {\tt AM1} & motion & 268, 268 & 0\\
		~ & {\tt AM1} & humidity & 459, 459  & 0\\
		~ & {\tt AM1} & luminance & 648, 648  & 0\\
		\hline
		\multirow{7}*{\tabincell{c}{T$_{3}$}}
		 & {\tt WS1} & water & 12, 12 & 0\\
		~ & {\tt MO6} & temperature &  6919, 6919 & 0\\
		~ & {\tt WL1} & water & 4, 4 & 0\\
		~ & {\tt HM1} & luminance & 1547, 1547 & 0 \\
		~ & {\tt HM2} & motion & 4046, 4046 & 0\\
		~ & {\tt HB1} & switch & 26,26  & 0\\
		~ & {\tt LB1} & switch & 18,18  & 0\\
		\hline
		\multirow{5}*{\tabincell{c}{T$_{4}$}}
		~ & {\tt MO7} & motion & 2116, 2116  & 0\\
		~ & {\tt HB4} & switch & 19, 19  & 0\\
		~ & {\tt LB2} & switch & 26, 26  & 0\\
		~ & {\tt HS1} & button & 15, 15 & 0\\
		~ & {\tt DL1} & level & 32 , 32  & 0\\
	\hline
	\end{tabular}
	\label{table_mediating_result}
	\vspace{-10pt}
\end{table}

\subsection{Performance of Data Filtering}
To evaluate the performance of our policy-based data filter, we run the four testbeds for another week with data filtering, i.e., we enable the execution of data-minimization policies. 
Also, two user-specified policies: UP1 (DO NOT report {\tt MO1} motion data between 5pm to 8am) and UP2 (DO NOT report {\tt MU1} contact data between 10am to 6pm) are defined in T$_{1}$.

\begin{table}[tb]
	\centering
	\scriptsize
	\caption{Results of correctness verification of all device commands that are issued by automation rules. $\rm N_{R}$: the number of device commands received by \tool after removing ones caused by the annotated manual operations; $\rm N_{S}$: the number of commands that are proved \emph{sound}; \mr{$\rm N_{C}$: the number of ground truth commands generated based on raw event replaying}. $\rm\mathbf{R_{S}}$ and $\rm\mathbf{R_{C}}$ are the ratio of commands that are sound and complete, respectively. 
	}
	\renewcommand\arraystretch{1}
	\newcommand{\tabincell}[2]{\begin{tabular}{@{}#1@{}}#2\end{tabular}}
	\newcolumntype{P}[1]{>{\raggedright\arraybackslash}p{#1}}
	\newcolumntype{M}[1]{>{\centering\arraybackslash}m{#1}}
	\newcolumntype{R}[1]{>{\raggedleft\arraybackslash}p{#1}}
    \begin{tabular}{|M{0.7cm}|M{0.6cm} |M{0.76cm}|M{0.5cm}|M{0.2cm}|M{1cm}| M{0.5cm}|M{1cm}|}
	\hline
		\multirow{2}{*}{\tabincell{c}{\textbf{Testbed}}} & \multirow{2}{*}{\tabincell{c}{\textbf{Device}}} & \multirow{2}{*}{\tabincell{c}{\bf Comm.}} & \multirow{2}{*}{\tabincell{c}{$\rm N_{R}$}}
		 & \multicolumn{2}{c|}{\textbf{Soundness}} & \multicolumn{2}{c|}{\textbf{Completeness}}  \\
		\cline{5-8}
		& & & & \tabincell{c}{$\rm N_{S}$} & \textbf{\tabincell{c}{$\rm \mathbf{R_{S}}$}}  & \tabincell{c}{$\rm N_{C}$} & \textbf{\tabincell{c}{$\rm \mathbf{R_{C}}$}}\\ \hline
		
		\multirow{4}*{\tabincell{c}{T$_{1}$}} 
		  & \tt OL1  & ON     & \bf 6(5) & 5 & \bf 0.83(1.00)   & 5  & 1.00 \\
		~ & \tt OL2  & ON     & 7 & 7 & 1.00   & 7  & 1.00 \\
		~ & \tt SL1  & ON     & 9 & 9 & 1.00   & 9  & 1.00  \\
		~ & \tt SL1  & OFF    & 9 & 9 & 1.00   & 9  & 1.00  \\
		\hline
		\multirow{9}*{\tabincell{c}{T$_{2}$}} 
		  & \tt OL3  & ON      & \bf 4(3)  & 3   & \bf 0.75(1.00) & 3  & 1.00  \\ %
		~ & \tt OL4  & ON      & \bf 21(21) & 20  & \bf 0.95(0.95) & 20 & 1.00  \\  %
		~ & \tt OL4  & OFF     & \bf 21(19) & 17  & \bf 0.81(0.89) & 17 & 1.00  \\  %
		~ & \tt SL2  & ON      & 10 & 10  & 1.00 & 10 & 1.00   \\ %
		~ & \tt SL3  & ON      & 31 & 31  & 1.00 & 31 & 1.00    \\ %
		~ & \tt SL4  & ON      & 20 & 20  & 1.00 & 20 & 1.00    \\ %
		~ & \tt SL4  & OFF     & 19 & 19  & 1.00 & \bf 20(19) & \bf 0.95(1.00)   \\  %
		~ & \tt SL5  & ON      & 13 & 13  & 1.00 & 13 & 1.00    \\   %
		~ & \tt SL5  & OFF     & 13 & 13  & 1.00 & 13 & 1.00    \\	%
		\hline
		\multirow{11}*{\tabincell{c}{T$_{3}$}} 
		  & \tt OL5  & ON                  &  7   & 7   &  1.00  &  7  & 1.00    \\ %
		~ & \tt OL5  & OFF                 &  7   & 7   &  1.00  &  7  & 1.00  \\ %
		~ & \tt OL6  & ON                  &  2   & 2   &  1.00  &  2  & 1.00  \\ %
		~ & \tt OL7  & ON                  &  9   & 9   &  1.00  &  9  & 1.00    \\ %
		~ & \tt SW1  & ON                  &  5   & 5   &  1.00  &  5  & 1.00   \\ %
		~ & \tt SW2  & ON                  &  23  & 23  &  1.00  &  23 & 1.00   \\ %
		~ & \tt SW2  & OFF                 &  \bf 19(18)  & 17  &  \bf 0.89(0.94)  &  17 & 1.00   \\ %
		~ & \tt SW3  & ON                  &  7   & 7   &  1.00  &  7  & 1.00   \\ %
		~ & \tt HB1  & ON                  &  7   & 7   &  1.00  &  7  & 1.00    \\ 
		~ & \tt HB2  & ON                  &  7   & 7   &  1.00  &  7  & 1.00   \\ 
		~ & \tt LB1  & ON                  &  9   & 9   &  1.00  &  7  & 1.00     \\ %
		\hline
		\multirow{14}*{\tabincell{c}{T$_{4}$}} 
		  & \tt OL8  & ON                  &  27  &  27 & 1.00 &  27 & 1.00     \\ %
		~ & \tt OL8  & OFF                 &  23  &  23 & 1.00 &  23 & 1.00   \\ %
		~ & \tt SW4  & ON                  &  67  &  67 & 1.00 &  67 & 1.00     \\ %
		~ & \tt SW4  & OFF                 &  55  &  55 & 1.00 &  55 & 1.00     \\ %
		~ & \tt HB3  & ON                  &  38  &  38 & 1.00 &  38 & 1.00    \\ 
		~ & \tt HB3  & OFF                 &  38  &  38 & 1.00 &  38 & 1.00     \\ %
		~ & \tt HB4  & ON                  &  34  &  34 & 1.00 &  34 & 1.00    \\ %
		~ & \tt HB4  & OFF                 &  34  &  34 & 1.00 &  34 & 1.00    \\ %
		~ & \tt HB5  & ON                  &  66  &  66 & 1.00 &  66 & 1.00     \\ %
		~ & \tt HB5  & OFF                 &  \bf 67(61) &  61 & \bf 0.91(1.00) &  61 & 1.00     \\ %
		~ & \tt LB2  & ON                  &  37  &  37 & 1.00 &  37 & 1.00     \\ %
		~ & \tt LB2  & OFF                 &  37  &  37 & 1.00 &  37 & 1.00     \\ %
		~ & \tt DL1  & Set 100             &  20  &  20 & 1.00 &  20 & 1.00     \\ %
		~ & \tt DL1  & Set 0               &  20  &  20 & 1.00 &  20 & 1.00     \\ %
	\hline
	\end{tabular}
	\label{table_policy_result}
\end{table}

\subsubsection{Correctness and Reliability}
\label{label_correctness_policy}
Comparing the received event sequences becomes meaningless since data are filtered in this setting. Instead, we evaluate the correctness of executions of automation rules. 

\vspace{2pt}\noindent\textbf{Methodology.}
\tool records all commands sent by the platform in each testbed. We ask the testbed members to keep a record of their manual operations on mobile companion apps and manually remove the commands caused by these operations; thus, the remaining commands are supposed to be issued by executions of rules; \mr{We use \emph{p-commands} to denote the commands that are issued by the platform when \tool is used.} \mr{We check if \emph{p-commands}} \mr{(i.e., the execution result from filtered events)} are \textbf{sound} (i.e., policies do not cause \mr{false positive} device actuation) and \textbf{complete} (i.e., policies do not lead to missed device actuation)
\mr{against \emph{ground truth} commands, denoted as \emph{gt-commands}, which are supposed to be issued by the platform using the same automation rules over the raw (unfiltered) events. To generate the gt-commands, we disconnect \tool from physical devices and replay the one-week raw events to the platform in each testbed, i.e., the raw events are sent to the platform over another week based on their timestamps. Note that all rules in Table~\ref{table_rules} are date-insensitive so that the dates of the timestamps of events does not affect the rules. For convenient comparison, we adjust the start time of gt-commands and make it the same as that of p-commands, i.e., make the two sets of commands start at the same time such that we can compare them. }

To verify the \emph{soundness} of each command in p-commands, we check if it has a counterpart within 3 seconds in the gt-commands. If so, the command is sound. Recall that \tool removes redundant rule executions (i.e., commands) while platforms not; for example, \tool will not send data for a rule to turn on a fan if the fan is already ``ON'' (see Section~\ref{section_ap}). Thus, the p-commands should not be considered \emph{incomplete} if it has no counterpart of a redundant command in the gt-commands. Therefore, we traverse the raw events and gt-commands in parallel and remove such redundant commands before verifying completeness. Then, we check if every command in the gt-commands has a counterpart within 3 seconds in the p-commands. 

\vspace{2pt}\noindent\textbf{Results.}
We compute the ratio $\rm \mathbf{R_{S}}$ and $\rm \mathbf{R_{C}}$ of commands that are sound and complete, respectively, as shown in Table~\ref{table_policy_result}. We find that 15 command records are not sound and 1 command (i.e., \texttt{SL4} OFF) in the gt-commands cannot find a match in the p-commands (marked in bold). Recall that we ask the testbed members to annotate their manual operations on mobile companion apps because these manual operations also generate commands which cannot pass the soundness verification. We ask the testbed members to recall and confirm if they miss annotating manual operations at the timestamp of the unsound commands. After rectifying the annotations, we repeat the soundness verification process. The updated results in the parentheses in Table~\ref{table_policy_result} show that $\rm \mathbf{R_{S}}$ increases after improving the user annotations. The remaining 4 unsound command records are very likely to be generated by 
forgotten manual operations. On the other hand, we look up the SmartThings live logging \cite{smartthings2020ide} and confirm that rule 13 (see Table~\ref{table_rules}) indeed issues an OFF command to device \texttt{SL4} at the timestamp of the unmatched command (\texttt{SL4} OFF). Therefore, the command may be lost during transmission from SmartThings cloud to \tool. The evaluation results prove that \tool does not impede the home automation.

\vspace{2pt}\noindent\textbf{Evaluation of a naive pull-based approach.}
\mr{Considering some platforms use both \emph{push} (i.e., devices report data to platforms via callbacks) and \emph{pull} (i.e., platforms request data from devices) models to obtain data, a \emph{naive} privacy-preserving approach is that \tool does not report data to the platform unless the platform proactively requests, on the assumption that the platform pulls data on demand. The SmartThings new system uses both push and pull. We use one of the SmartThings testbed T$_{3}$ to evaluate this baseline (pull-based) approach. We keep the configuration of devices and rules the same and modify the behavior of \tool, i.e., \tool responds to \texttt{State Refresh} (pull) with unfiltered real data but does not push any data. We run this new setting for a week and use the aforementioned correctness verification methodology to check if the baseline approach guarantees the correctness of automation.} 

\mr{
The results in Table~\ref{table_naive_result} show that the baseline approach leads to failures (non-executions) of most automation rules (i.e., rules in line 3-11) except rule 20 (line 1 \& 2; device: OL5), which only uses a specific time to trigger the rule. According to our observation, SmartThings only pulls the state of a specific device when we operate on the SmartThings mobile app to refresh its state; otherwise, SmartThings relies on the \emph{push} model to update devices' states and trigger rules. Thus, when the naive approach is employed, most automation rules cannot be triggered because the platform does not know the events that trigger the rules. 
} 




\begin{table}[tb]
	\centering
	\scriptsize
	\caption{\mr{Results of the naive pull-based approach. See Table~\ref{table_policy_result} for the notations. }
	}
	\renewcommand\arraystretch{1}
	\newcommand{\tabincell}[2]{\begin{tabular}{@{}#1@{}}#2\end{tabular}}
	\newcolumntype{P}[1]{>{\raggedright\arraybackslash}p{#1}}
	\newcolumntype{M}[1]{>{\centering\arraybackslash}m{#1}}
	\newcolumntype{R}[1]{>{\raggedleft\arraybackslash}p{#1}}
    \begin{tabular}{|M{0.7cm}|M{0.6cm} |M{0.76cm}|M{0.5cm}|M{0.2cm}|M{1cm}| M{0.5cm}|M{1cm}|}
	\hline
		\multirow{2}{*}{\tabincell{c}{\textbf{Testbed}}} & \multirow{2}{*}{\tabincell{c}{\textbf{Device}}} & \multirow{2}{*}{\tabincell{c}{\bf Comm.}} & \multirow{2}{*}{\tabincell{c}{$\rm N_{R}$}}
		 & \multicolumn{2}{c|}{\textbf{Soundness}} & \multicolumn{2}{c|}{\textbf{Completeness}}  \\
		\cline{5-8}
		& & & & \tabincell{c}{$\rm N_{S}$} & \textbf{\tabincell{c}{$\rm \mathbf{R_{S}}$}}  & \tabincell{c}{$\rm N_{C}$} & \textbf{\tabincell{c}{$\rm \mathbf{R_{C}}$}}\\ \hline
		
		\multirow{11}*{\tabincell{c}{T$_{3}$}} 
		  & \tt OL5  & ON                  &  7   & 7   &  1.00  &  7  & 1.00    \\ %
		~ & \tt OL5  & OFF                 &  7   & 7   &  1.00  &  7  & 1.00  \\ %
		~ & \tt OL6  & ON                  &  0   & 0   &  --  &  3  & 0.00  \\ %
		~ & \tt OL7  & ON                  &  0   & 0   &  --  &  8  & 0.00    \\ %
		~ & \tt SW1  & ON                  &  0   & 0   &  --  &  5  & 0.00   \\ %
		~ & \tt SW2  & ON                  &  0  & 0  &  --  &  27 & 0.00   \\ %
		~ & \tt SW2  & OFF                 &  0  & 0  &  --  &  23 & 0.00   \\ %
		~ & \tt SW3  & ON                  &  0   & 0   &  --  &  8  & 0.00   \\ %
		~ & \tt HB1  & ON                  &  0   & 0   &  --  &  7  & 0.00    \\ 
		~ & \tt HB2  & ON                  &  0   & 0   &  --  &  7  & 0.00   \\ 
		~ & \tt LB1  & ON                  &  0   & 0   &  --  &  7  & 0.00     \\ %
	\hline
	\end{tabular}
	\label{table_naive_result}
\end{table}

\subsubsection{Latency}
Compared to the original systems, \tool adds one more ``hop'' to the original system. Therefore, the additional latency $L_{\rm HA}$ introduced by \tool to home automation consists of the computation latency $L_{1}$ for executing policies and the additional transmission delay. The additional transmission delay includes the event transmission delay $L_{2}$ from \tool to the local hub (SmartThings classic) or to the platform (the new SmartThings, openHAB)
and the command transmission delay in the reverse direction. Thus, $L_{\rm HA}$ is approximately equal to the sum of computation latency $L_{1}$ and a round-trip transmission delay $2*L_{2}$, i.e., $L_{\rm HA}=L_{1}+2*L_{2}$. We obtain $L_{1}$ by computing the elapsed time from when \tool receives an event to when \tool reports the event (if the event is reported), and we obtain $L_{2}$ by computing the elapsed time from when \tool reports an event to when the hub/platform receives the same event.  We compute the latency for all events and show the averaged results in Table~\ref{table_latency}. The extra latency (about 0.6 second) is a tradeoff for mitigating privacy leakage. User experience suffers less than indicated by the latency since many rules (e.g., rule 2, 3, 4, 5 in Table~\ref{table_rules}) are not time-critical. Hence, the latency (up to 0.7 seconds) of most automation rules is acceptable.

\begin{table}
	\scriptsize
	\caption{Latency introduced by \tool (in seconds).}
	\renewcommand\arraystretch{1}
	\newcommand{\tabincell}[2]{\begin{tabular}{@{}#1@{}}#2\end{tabular}}
	\newcolumntype{P}[1]{>{\arraybackslash}p{#1}}
	\newcolumntype{M}[1]{>{\centering\arraybackslash}m{#1}}
	\begin{tabular}{M{0.8cm}M{1.8cm}M{2.8cm}M{1.5cm}}
		\toprule
		\bf \tabincell{c}{Testbed} &  \tabincell{c}{Computation\\Latency $L_{1}$} & \tabincell{c}{One-Way Transmission \\Latency $L_{2}$} & \tabincell{c}{Automation\\ Latency $L_{\rm HA}$} \\ \midrule
		T$_{1}$ &  0.136  &  0.233  &  0.602 \\
		T$_{2}$ &  0.087  &  0.203  &  0.493  \\
		T$_{3}$ &  0.143  &  0.274  &  0.691  \\
		T$_{4}$ &  0.083  &  0.293  &  0.669  \\
		\bottomrule
	\end{tabular}
	\label{table_latency}
\end{table}

\subsubsection{Reduction of Data Leakage}
\label{section_data_reduction}
To show the effectiveness of \tool in filtering events, we compare the data volume collected by \tool with that reported to the platforms and compute the relative Reduction Rate ($\rm RR$). As shown in Table~\ref{table_filtering_data_result}, \tool blocks more than 99\% of numeric-value sensor readings, and 96.79\% of binary-value sensor readings and device states (i.e., motion, presence, switch, etc.) which reveals sensitive information. By reducing the disclosed data, \tool can effectively prevent smart home platforms and potential attackers from inferring private information of smart homes and homeowners based on large amount of IoT device data in the original system. In general, the $\rm RR$ of binary-value attributes are smaller than numeric-value ones, since binary attributes are more often used as rule triggers and hence cannot be completely blocked. 

\begin{table}[tb]
	\centering
	\scriptsize
	\caption{Results of the reduction rate $\rm RR$ and correct tracking ratio $\rm CTR$/$\rm CATR$ per device. $\rm VOL$: the volume of raw event and the volume of event reported to the platform after data filtering, respectively. The last column is $\rm CATR$ if not specified and otherwise $\rm CTR$. We present the results for partial representative devices due to page limits.
	}
	\renewcommand\arraystretch{1}
	\newcommand{\tabincell}[2]{\begin{tabular}{@{}#1@{}}#2\end{tabular}}
	\newcolumntype{P}[1]{>{\raggedright\arraybackslash}p{#1}}
	\newcolumntype{M}[1]{>{\centering\arraybackslash}m{#1}}
	\newcolumntype{R}[1]{>{\raggedleft\arraybackslash}p{#1}}
	\begin{tabular}{|M{0.8cm}|M{0.6cm}|M{1.2cm}|M{1.0cm}|M{0.6cm}|M{1.9cm}|}
	\hline
		\bf Testbed & \centering\bf Device  & \bf Attribute  & $\rm VOL$ & $\rm\mathbf{RR}$ & \bf $\rm\mathbf{CATR}$ ($\rm\mathbf{CTR}$) \\\hline	 
		\multirow{6}*{\tabincell{c}{T$_{1}$}} 
		& {\tt MU1} & contact & 1244, 9 & 0.99 & 0.01 \\
		~ & {\tt MO1} & motion & 1574, 26 & 0.98 & 0.75 \\
		~ & {\tt OL1} & switch & 22, 4 & 0.82 & 0.32 \\
		~ & {\tt SL1} & switch & 18, 18 & 0.00 & 1.00 \\
		~ & {\tt PR1} & presence & 42, 30 & 0.29 & 0.91 \\
		~ & {\tt PR2} & presence & 42, 1 & 0.99 & 0.32 \\
		\hline
		\multirow{6}*{\tabincell{c}{T$_{2}$}}
		  & {\tt MU2} & contact & 30, 10    & 0.67 & 0.01 \\
		~ & {\tt MO2} & motion  & 264, 45   & 0.83 & 0.46 \\
		~ & {\tt PR6} & presence & 22, 4    & 0.82 & 0.48 \\
		~ & {\tt OL3} & switch & 36, 4      & 0.89 & 0.03 \\
		~ & {\tt AM1} & illumance & 658, 1  & 0.99 & 0.00 ($\rm CTR$) \\
		~ & {\tt AM1} & humidity & 487, 0   & 1.00 & 0.00 ($\rm CTR$) \\
		\hline
		\multirow{7}*{\tabincell{c}{T$_{3}$}}
		 & {\tt WS1} & water & 14, 5 & 0.64 & 0.03 \\
		~ & {\tt MO4} & motion & 918, 45& 0.95 & 0.15 \\
		~ & {\tt MO6} & motion & 923, 7 & 0.99 & 0.38 \\
		~ & {\tt HM1} & motion & 1090, 2 & 0.99 & 0.05 \\
		~ & {\tt HM2} & motion & 5046, 2 & 0.99 & 0.31 \\
		~ & {\tt HB1} & switch & 24, 12 & 0.50 & 0.22 \\
		~ & {\tt WL1} & water & 4, 0 & 1.00 & 0.00 \\
		\hline
		\multirow{7}*{\tabincell{c}{T$_{4}$}}
		~ & {\tt HM3} & motion & 1354, 122 & 0.91 & 0.27 \\
		~ & {\tt HM4} & motion & 1178, 50 & 0.96 & 0.55 \\
		~ & {\tt HM5} & motion & 5778, 68 & 0.99 & 0.48 \\
		~ & {\tt HM6} & motion & 2815, 127 & 0.95 & 0.22 \\
		~ & {\tt MO7} & motion & 1898, 74 & 0.96 & 0.51 \\
		~ & {\tt HB3} & switch & 338, 122 & 0.64 & 0.79 \\
		~ & {\tt HS1} & button  & 16, 14  & 0.13 & 0.54 \\
	\hline
	\end{tabular}
	\label{table_filtering_data_result}
	\vspace{-10pt}
\end{table}

\subsubsection{Privacy Gain of \tool}
\label{privacy_gain}
We further investigate how much privacy \tool enhances by studying the degree that \tool prevents attackers from: (1) tracking device states, and (2) inferring sensitive user activities.

\vspace{3pt}\noindent
\textbf{Device Status Tracking.} An attacker can precisely track the state of a device by leveraging its complete event sequence. When \tool is deployed, it filters out a very large portion of device events such that it is not easy for an attacker to infer the correct device state with partial information. We define two metrics for this study: Correct Tracking Ratio (CTR) and Correct Active-State Tracking Ratio (CATR):

\vspace{-1em}
\begin{equation}
\begin{aligned}
\footnotesize
\rm CTR = \frac{\text{Duration of correctly guessing a sensor's measurements}}{\text{Total duration}}
\end{aligned}
\vspace{-10pt}
\end{equation}

\begin{equation}
\begin{aligned}
\footnotesize
\rm CATR = \frac{\text{Duration of correctly guessing a device's active state}}{\text{Duration of guessing a device's active state}}
\end{aligned}
\end{equation}

CTR measures the correctness of tracking numeric-value IoT attributes (e.g., temperature, humidity) and CATR measures the correctness of tracking binary-value attributes (e.g., motion and switch). CTR is not suitable for measuring binary-value attributes; for instance, a CTR is equal to 80\% when simply guessing a motion sensor (which is ``inactive'' for 80\% of a day) is ``inactive'' all day. However, this does not indicate a good inference since an active state (e.g., motion ``active'', switch ``on'') matters more than the other (e.g., motion ``inactive'', switch ``off'') from the perspective of activity inference. Instead, CATR only takes the active state of a binary-value attribute into calculation and thus eliminates the impact of unbalanced distributions of binary states.

We calculate the CTR or CATR of every attribute for all devices in the four testbeds over one week. The result is shown in Table~\ref{table_filtering_data_result}. The CTR of all numeric-value attributes are equal or close to zero since most of them are not used by automation rules and hence are not reported to the platform. Although some numeric-value attributes are used in rule conditions (e.g., \texttt{AM1}-luminance used by rule 8), more than 99\% of their values are not reported and therefore attackers cannot correctly track them. Overall, most binary-value attributes have a low CATR, showing that data filtering prevents attackers from correctly tracking the device status. The CATR of a few binary-value attributes (e.g., \texttt{PR1}-presence) is higher because these attributes are heavily needed by rules and thus most of their values cannot be filtered out.

\vspace{3pt}\noindent
\textbf{Activity Inference.}
We consider a strong attacker who has sufficient background and domain knowledge to perform sophisticated attacks to infer user activities. Smart home users usually give a meaningful name (e.g., bathroom motion sensor, bedroom light, microwave outlet) and assign a room to each device. Hence, we assume that the attacker knows the rooms of a home and the assigned room of every device. We also assume that the attacker has expertise to associate activities with device events and states.\footnote{Such knowledge can be easily learned from a lot of existing literature such as \cite{rashidi2010discovering, hammid2012unsupervised, gu2009epsicar, viard2016event}.} Based on the above assumptions of a strong attacker, we implement methods (see Table~\ref{table_activity_inference}) of attackers to infer users' activities. We run the methods on raw events (before data filtering) and the processed events (after data filtering) and compare the results.

\begin{figure}[t]
  \centering
  \subfigure[Without data filtering]{
    \label{fig_presence_without} 
    \includegraphics[width=0.23\textwidth]{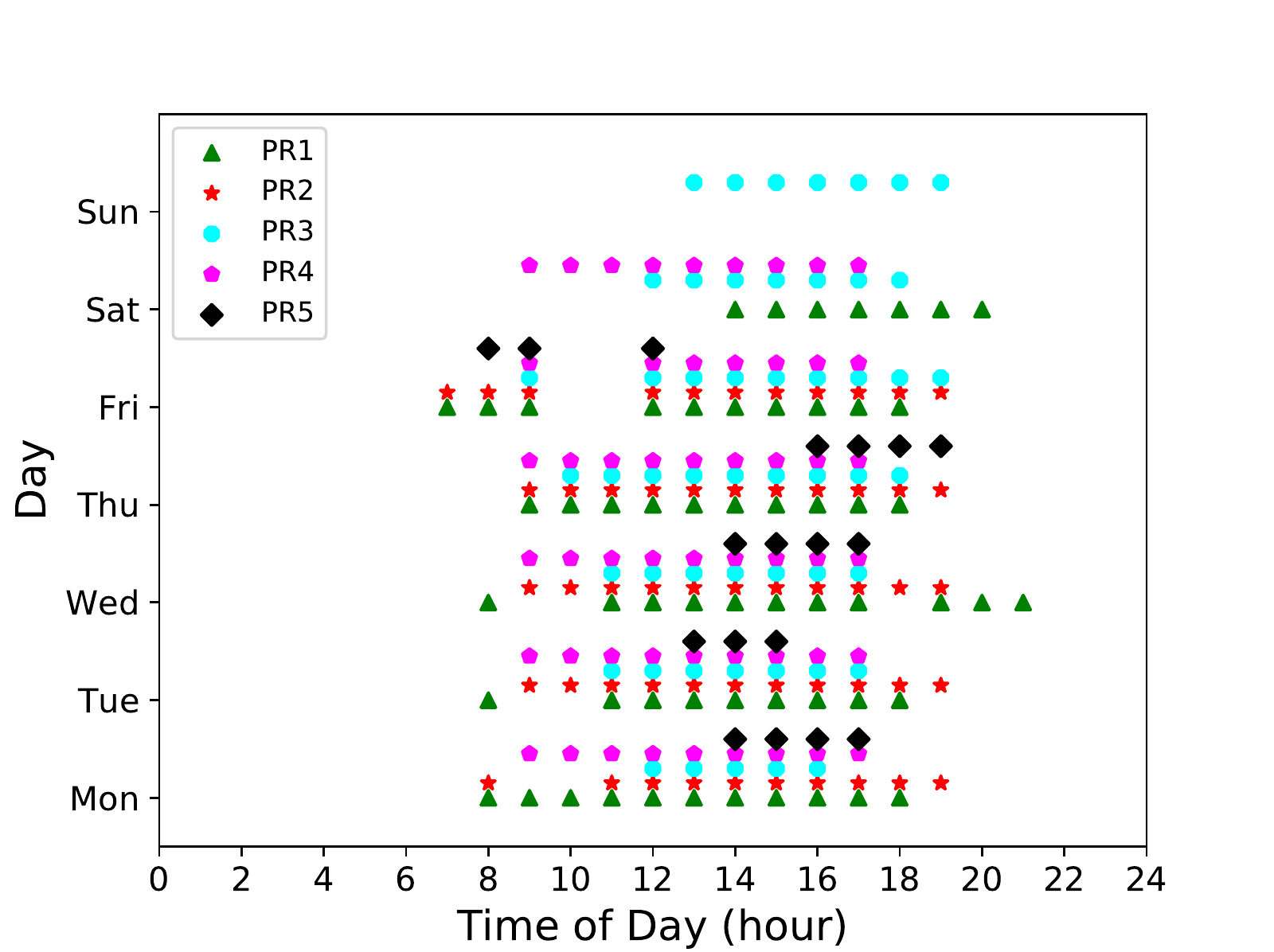}}
  \subfigure[With data filtering]{
    \label{fig_presence_with}
    \includegraphics[width=0.23\textwidth]{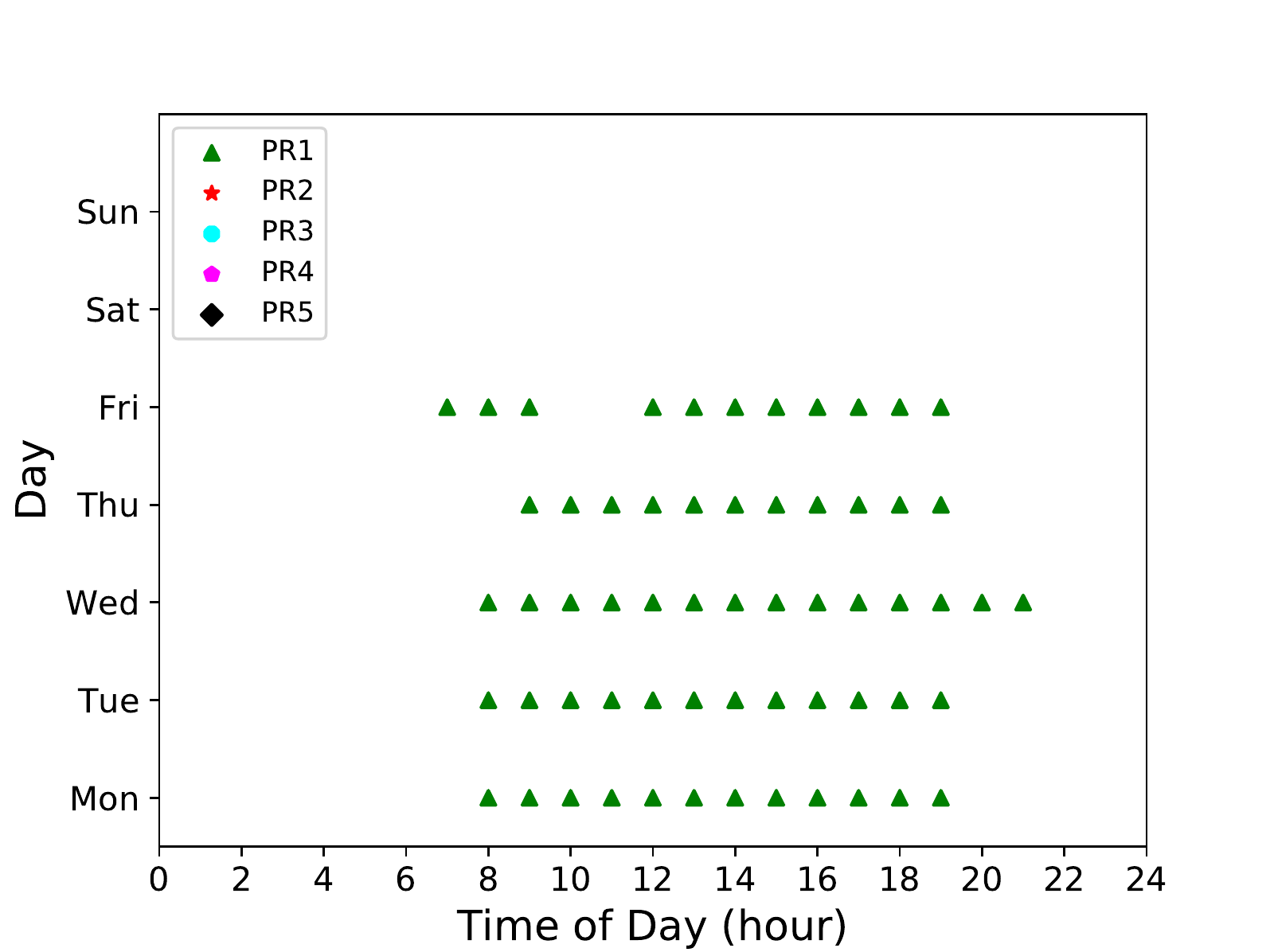}}
  \caption{Visualization of inferred working hours based on the presence sensor data in testbed T$_{1}$ without and with data filtering, respectively. For simplicity of illustration, we round all working hours to the nearest hours. \texttt{PR1}$\sim$\texttt{5}: presence sensors in T$_{1}$.}
  \label{fig_working_hour_examples}
\end{figure}

As shown in Fig~\ref{fig_working_hour_examples}, the working hours of every member who carries a presence sensor can be easily inferred by observing the raw presence sensor data. Note that \texttt{PR1}$\sim$\texttt{5} are the 5 presence sensors in testbed T$_{1}$. Furthermore, the device label of a presence sensor may carry personally identifiable information (e.g., name) and thus the learned working hours could be matched to individuals. When \tool is deployed, most presence data are filtered out and thus attackers cannot infer working hours correctly from \texttt{PR2}$\sim$\texttt{5}. Attackers can still achieve an accuracy of since the data of \texttt{PR1} cannot be completely filtered since both the states of \texttt{PR1} (i.e., ``present'' and ``not present'') are needed for executing automation rules 2 and 3 (see Table~\ref{table_rules}). 

\begin{figure}[t]
  \centering
  \subfigure[T$_{3}$ without data filtering]{
    \includegraphics[width=0.23\textwidth]{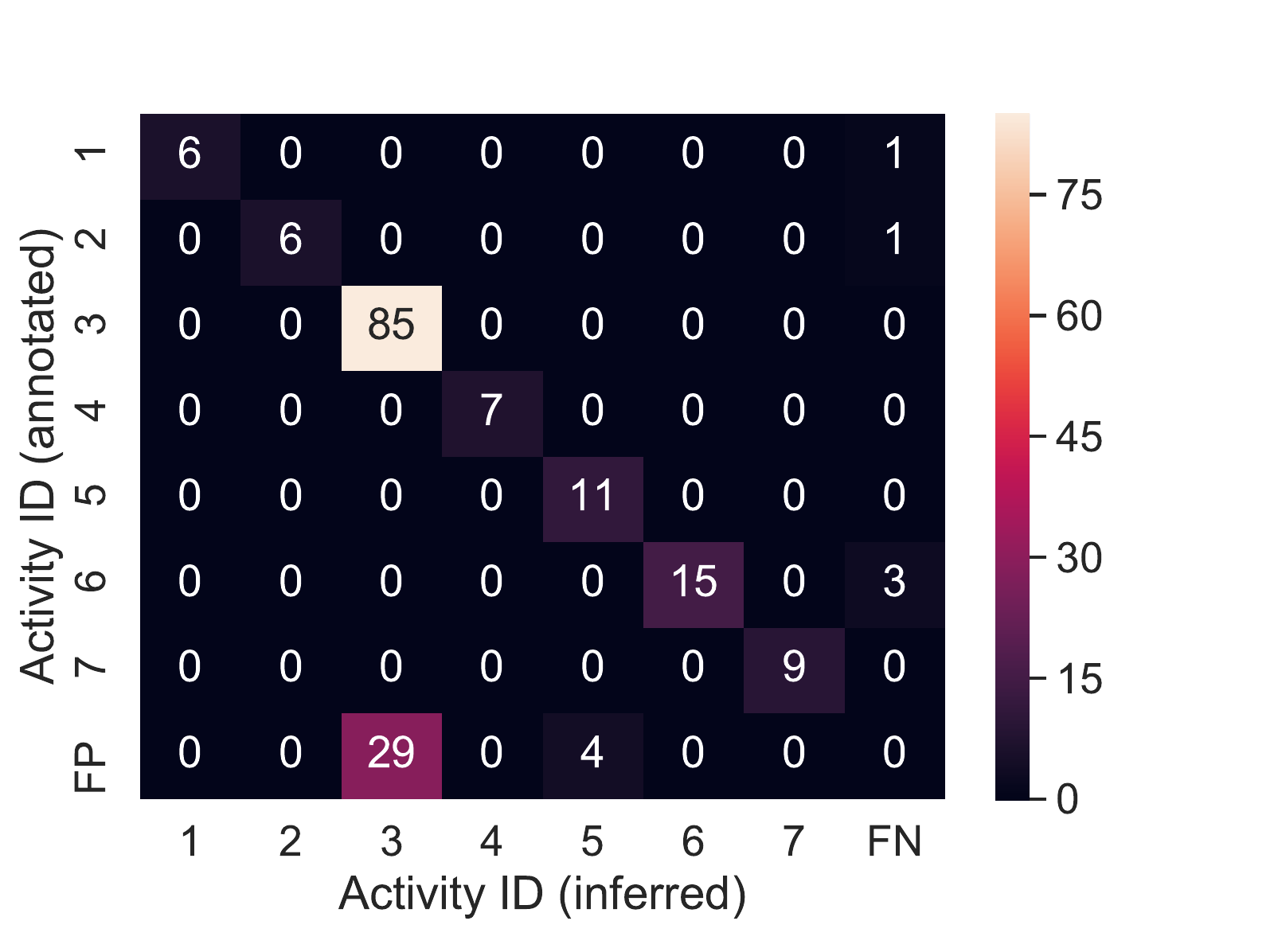}}
  \subfigure[T$_{3}$ with data filtering]{
    \includegraphics[width=0.23\textwidth]{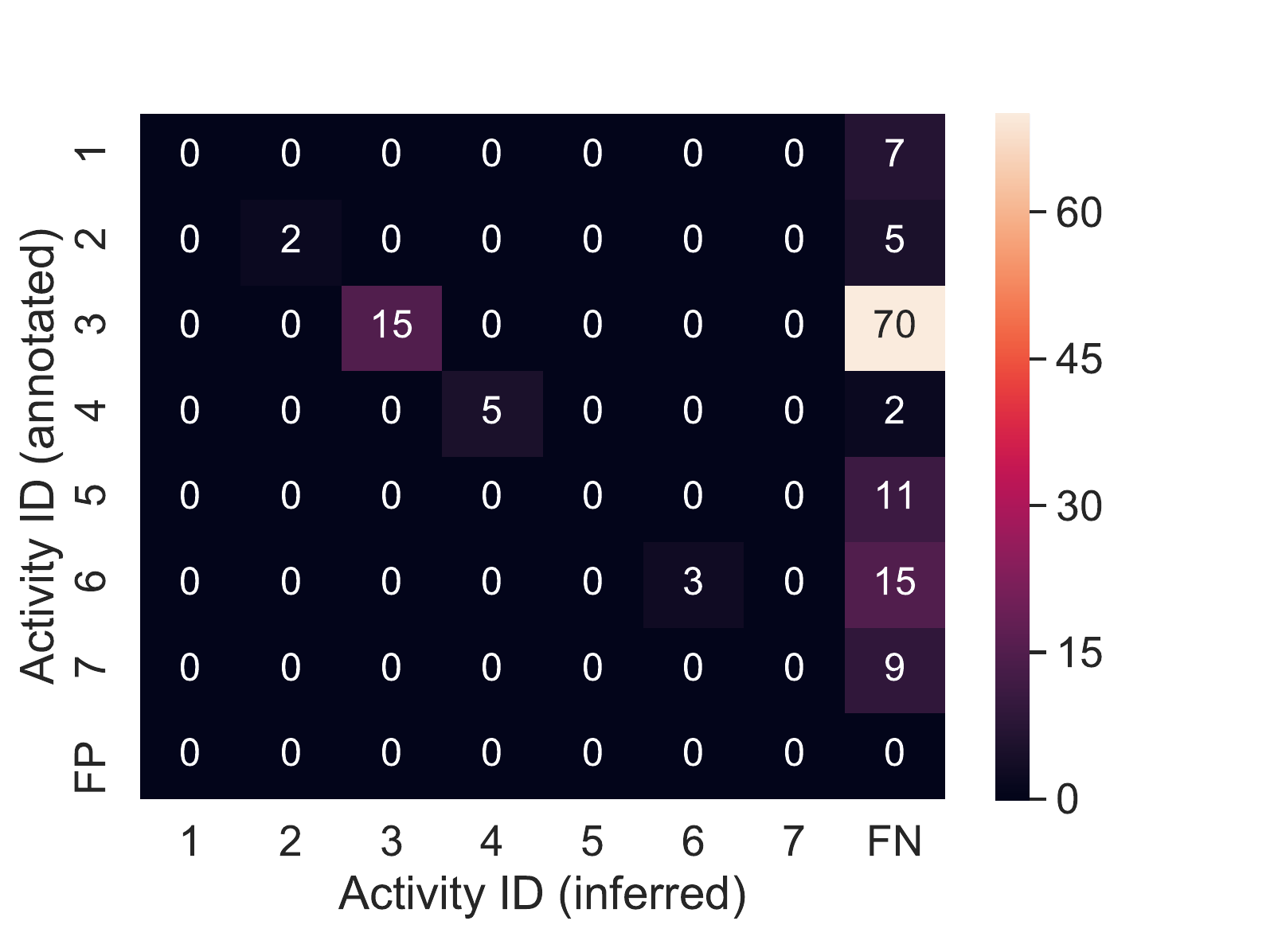}}
  \subfigure[T$_{4}$ without data filtering]{
    \includegraphics[width=0.23\textwidth]{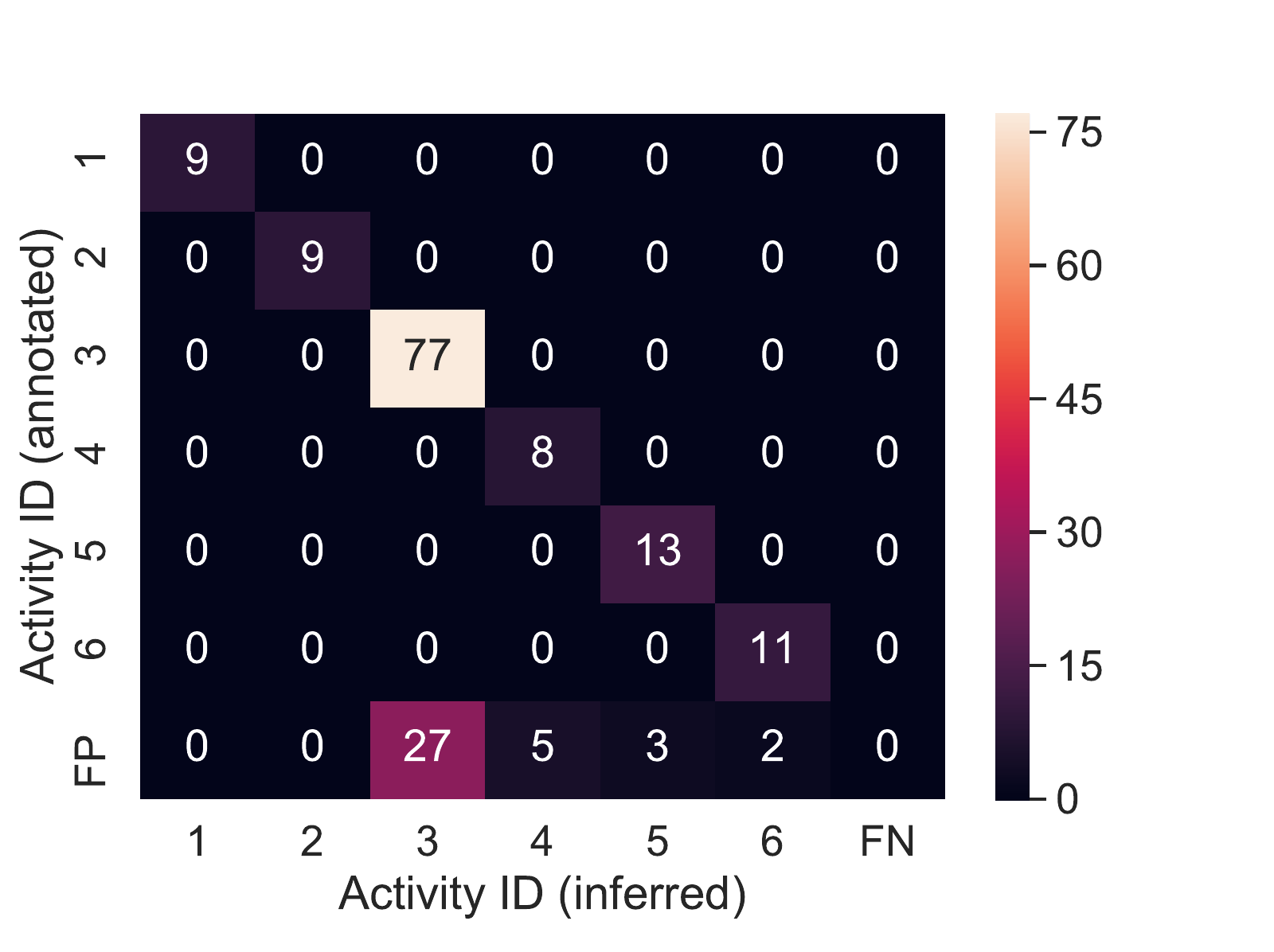}}
  \subfigure[T$_{4}$ with data filtering]{
    \includegraphics[width=0.23\textwidth]{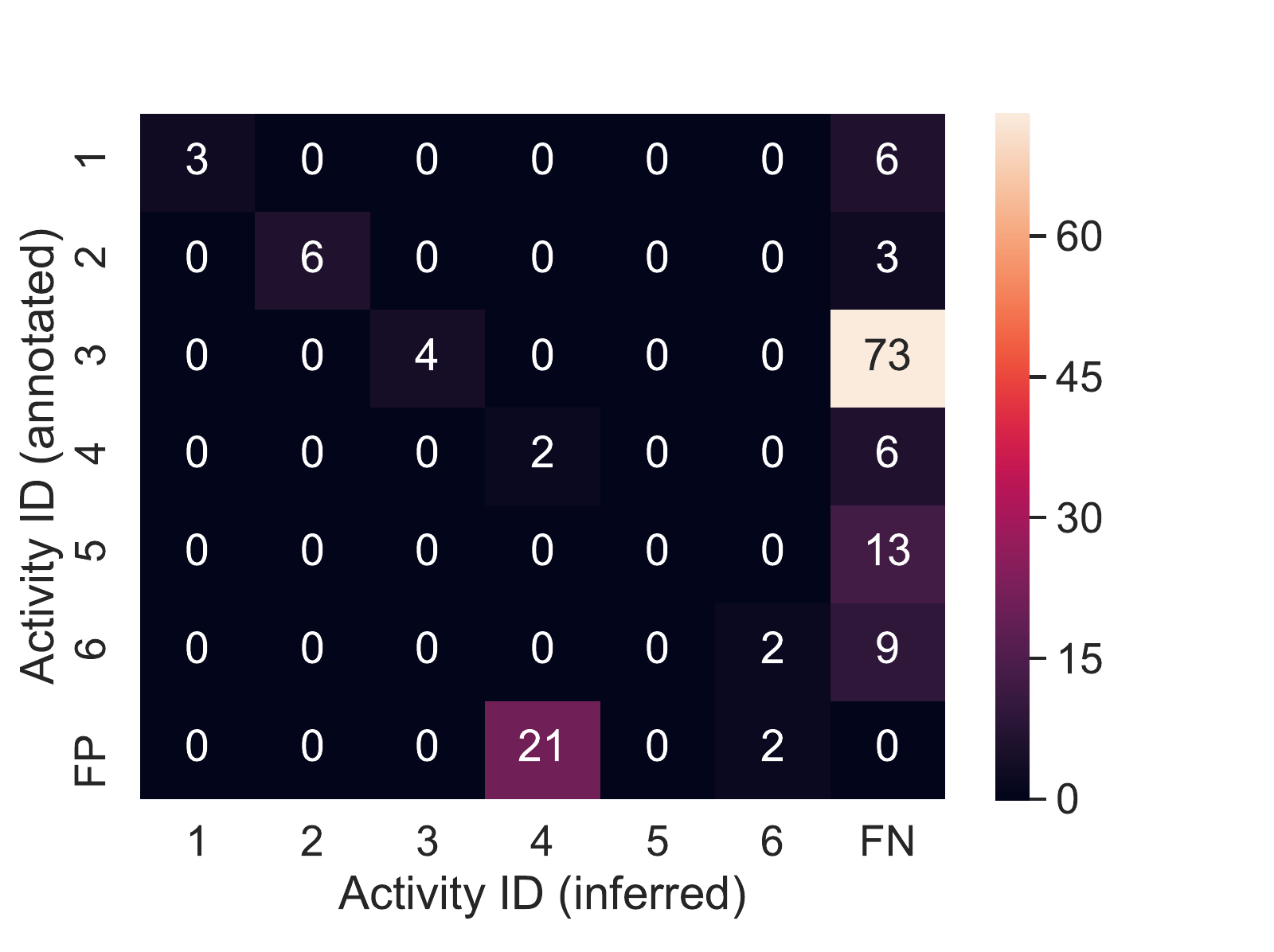}}
  \caption{Results of activity inference attacks (see Table~\ref{table_activity_inference}) in testbed T$_{3}$ and T$_{4}$. FN and FP denote false negative and false positive inference, respectively.}
  \label{fig_AR_results}
\end{figure}

Figure~\ref{fig_AR_results} shows the confusion matrices of activity inference attacks before and after data filtering on testbed T$_{3}$ and T$_{4}$. Without \tool, the attacker correctly recognizes most activities. The \emph{recall} (a.k.a., true positive rate) is 1 for most activities except that it is 0.83 for activities 1, 2, 6 in T$_{3}$. Data filtering brings the recall of most activities down to less than 0.34 and some (activity 1, 5, 7 in T$_{3}$ and 5 in T$_{4}$) even to 0, making the attacker miss recognizing most activities. Overall, the total recall of all activities drops from 0.96 to 0.19 in T$_{3}$ and from 1 to 0.13 in T$_{4}$. Hence, by filtering data, \tool significantly degrades the performance of the activity inference attacks.

\begin{table}[tb]
	\centering
	\scriptsize
	\caption{Methodology for inferring user activities in different testbeds. Device, room and time information is exploited.
	}
	\renewcommand\arraystretch{1.1}
	\newcommand{\tabincell}[2]{\begin{tabular}{@{}#1@{}}#2\end{tabular}}
	\newcolumntype{P}[1]{>{\raggedright\arraybackslash}p{#1}}
	\newcolumntype{M}[1]{>{\centering\arraybackslash}m{#1}}
	\newcolumntype{R}[1]{>{\raggedleft\arraybackslash}p{#1}}
	\begin{tabular}{|M{0.9cm}|M{0.2cm}|P{5.2cm}|M{0.7cm}|}	
	\hline
	    \textbf{Activity} & \textbf{ID}             & \centering\textbf{Methodology} & \textbf{Testbed} \\ \hline
	    Working & 0    & presence sensor on  & T$_{1}$ \\ \hline
	    Leaving home & 1  & \tabincell{l}{front door open/closed$\rightarrow$all motion sensors become\\ inactive for at least 10 minutes}  & \tabincell{c}{T$_{3}$, T$_{4}$}  \\ \hline
	    Arriving home & 2  & \tabincell{l}{front door open/closed $\rightarrow$ at least a motion sensor \\ becomes active within 3 minutes}  & \tabincell{c}{T$_{3}$, T$_{4}$} \\ \hline
	    Toileting  & 3                & 1 minute $<$ bathroom motion active $<$ 10 minutes & \tabincell{c}{T$_{3}$, T$_{4}$} \\ \hline
	    \multirow{2}{*}{Showering} & \multirow{2}{*}{4} & bathroom motion active $\rightarrow$ water sensor wet & T$_{3}$   \\ \cline{3-4}
	    ~          &                & 15 minutes$<$bathroom motion active$<$60 minutes & T$_{4}$    \\ \hline
	    Sleeping   & 5                & \tabincell{l}{bedroom motion active$\rightarrow$no motion active in other\\ rooms$\rightarrow$bedroom motion inactive$>$10 minutes} & \tabincell{c}{T$_{3}$, T$_{4}$}  \\ \hline
	    \multirow{2}{*}{Cooking} & \multirow{2}{*}{6}   & kitchen motion $>$ 10 minutes & \tabincell{c}{T$_{3}$, T$_{4}$}  \\ \cline{3-4}
	    ~           &               & microwave outlet power $>$ 1000W & T$_{3}$  \\ \hline
	    Preparing coffee &  7  &  coffee machine outlet power $>$ 1000W & T$_{3}$  \\ \hline
	\end{tabular}
	\label{table_activity_inference}
	\vspace{5pt}
\end{table}

\section{Related Work}
\label{section_related_work}

\subsection{Smart Home Security}
Recent research have explored smart home security and safety in various aspects. 
Fernandes et al. \cite{fernandes2016security} revealed design vulnerabilities in the permission system of SmartThings. Follow-up works \cite{jia2017contexiot, tian2017smartauth, rahmati2018tyche, lee2017fact, demetriou2017hanguard} have designed new mechanisms to improve access control in IoT systems.
Security researches also studied the application security. For instance, Jia et al. \cite{jia2017contexiot} extensively surveyed app-level attacks on IoTs, and presented a context-based permission system. \fv{Other than single-app attacks, cross-app interaction (CAI) threats attracted the attention of recent works \cite{chi2018cross, chi2020cross, celik2018soteria, nguyen2018iotsan, celikiotguard}.}
\zz{HomeGuard~\cite{chi2018cross, chi2020cross} is the first work that systematically
studies this problem and also the first one that proposes to use theorem proving to
detect CAI threats. iRuler~\cite{wang2019charting} applies the same approach to studying
CAI threats on the IFTTT platform.}

\zz{In addition to automation applications, resource-constrained and diverse IoT devices raise unique challenges. For example, how to conduct IoT authentication~\cite{li2019touch} and pairing~\cite{li2020t2pair} for IoT devices that only have a button or knob is an 
intriguing question. As another example,
researchers frequently employ data-driven techniques to detect malfunctions in smart homes but the detection is hardly precise and explainable. HAWatcher \cite{fu2021hawatcher} proposes a novel semantics-aware approach that utilizes semantic information, such as automation rules and device relations, to mine correlations from event logs for precise and explainable anomaly and attack detection. Security
issues related to voice assistants draw great attention~\cite{carlini2016hidden, carlini2018audio, zeng2019multiversion}. For example, MVP-Ears~\cite{zeng2019multiversion} presents a highly accurate (accuracy $>$ 99.8\%)
audio adversarial example (AE) detection method inspired by multiversion programming, and it can proactively handle transferable audio AEs. This work is concerned with the rich
privacy-sensitive data generated by IoT devices.}

\subsection{Smart Home Privacy}
Privacy is an important research topic in smart home ecosystems. Zheng et al. \cite{zheng2018user} studied smart home owners' perceptions of privacy risks and actions taken to protect their privacy. The study found that users are unaware of privacy risks from inference algorithms operating on data from their IoT devices, and they expect device manufacturers to protect their privacy though it is not the case. Celik et al. \cite{celik18sensitive} tracked the sensitive data flows in programming frameworks and identified that 138 out of 230 apps in SmartThings transmit at least one kind of sensitive data over platform-provided APIs, which means malicious apps have the capability to steal user data collected by the platform. The literature \cite{fernandes2016security, jia2017contexiot}, \mr{\cite{surbatovich2017some, manandhar2020towards}} also demonstrated that \mr{IoT apps/rules can be exploited to breach user privacy}. FlowFence \cite{fernandes2016flowfence} enforced a data flow control mechanism for sensitive data protection. However,
unlike our work, FlowFence protects sensitive data from unauthorized apps rather than the platform, which is trusted in their threat model. 
Plus, FlowFence requires the cooperation from the platform and app developers to operate.
\zz{Xu et al.~\cite{xu2019privacy} filter and obfuscate data sent from IoT platforms 
to IFTTT to enhance smart home privacy;
however, the user data still flow to IoT platforms. How to protect smart
home privacy from such IoT platforms as SmartThings and Amazon has not been studied
prior to our work.}

\subsection{In-hub Security and Privacy Enforcement}
Many in-hub schemes are proposed to enforce security and privacy schemes in the IoT domain. Simpson et al. designed an in-hub security manager built atop the smart home hub to patch vulnerable IoT devices and strengthen authentication \cite{simpson2017securing}. The security manager is deployed in an open-source system HomeOS. FACT \cite{lee2017fact} and HanGuard \cite{demetriou2017hanguard} enforced access controls in the middle by implementing controllers on an open-source hub and a programmable WiFi router, respectively. By comparison, these schemes rely on a programmable hub (gateway, router) that can indeed intercept and control the communication between the home area network and the Internet. However, in cloud-based smart home platforms like SmartThings, communications between the commercial hub and the backend cloud are encrypted \cite{veracodehubresearch} and hence the router can neither decrypt nor modify the packets on demand. \tool provides
a novel approach to working with closed systems, without modifying the IoT devices, hubs, or clouds.


\subsection{Firewall-Based Solutions}
\mr{
Network firewalls have been extensively studied in different communication layers, such as data-link \cite{khan1997design, liu1999packet}, network \cite{park2001effectiveness, ingham2002history}, transport \cite{cheswick1990design}, and application \cite{viegas2016towards, krueger2010tokdoc, prandl2015study}, and in different aspects, such as policy compactness \cite{yoon2009minimizing, bodei2018language}, verification \cite{youssef2009automatic}, languages \cite{zhang2007specifications}, conflicts \cite{al2003firewall, maldonado2015detection}, and so on. \tool is essentially an application layer filtering system. Existing application firewalls, such as 
intrusion prevention systems \cite{viegas2016towards} and web application firewalls \cite{krueger2010tokdoc, prandl2015study}, check on non-payload information (e.g., IP address, port, protocol, DNS, HTTP URL, etc.). On the other hand,
\tool is a novel contextualized solution to the privacy protection of smart homes by inspecting application-layer payloads (i.e., events) and performing semantics-aware data filtering against the ``trigger-condition-action'' automation rules. 
}

\section{Discussion and Limitations}
\label{section_discussion_limitations}
\noindent

\vspace{3pt}
\noindent
\textbf{Can \tool perform home automation and thus get rid of
the cloud?}
Theoretically, \tool is capable of executing the extracted rules locally and disclosing no data to the platform. However, we did not employ this design due to practical considerations. (1) The kick-cloud-out strategy may cause ethical or legal concerns which our research team cannot tackle. 
While \tool may provide home automation, all other cloud-based services (messaging, storage, and remote management) will be lost. (2) Huge engineering efforts are needed to implement an equivalent rule engine that supports the same programming framework and APIs as well as maintaining them in a long run. 
Therefore, we strategically separate the data-filtering policy engine and the rule engine; and \tool only deals with data filtering.

\vspace{3pt}\noindent\textbf{Pull Strategies.}
\mr{There are two pull-based models for home automation: \emph{batch pull} and \emph{lazy pull}. A representative platform employing \emph{batch pull} is IFTTT, which polls the most recent $N$ (50 by default) events about once every hour \cite{iftttapi}. This model still does not take data minimization into consideration. \emph{Lazy pull} is a naive privacy-preserving framework where platforms only pull the current state of devices when needed instead of expecting devices to continuously push data to them. \emph{Lazy pull} is probably impractical for home automation, as the platforms cannot predict the occurrence of device events. To our best knowledge, no smart home platforms use \emph{lazy pull}.}

\vspace{3pt}\noindent\textbf{User efforts.}
To deploy \tool, users only need to pair their IoT devices with \tool mediator and operate on the mobile companion app to connect \tool with the target platform. 
All these operations are similar to using their original platforms and users do not need any expertise.

\vspace{3pt}\noindent\textbf{Compatibility.}
\mr{
Besides home automation, a dashboard function for viewing and controlling device states is usually provided on a companion mobile app by the IoT platform. With \tool, the dashboard may not always display the correct device states. However, this does not undermine the value of \tool, which can be deployed as a privacy enhancing \emph{add-on} for privacy-conscious users who mainly use the platform for automation. It is worth noting that occasional remote access from the companion app does not justify that IoT data should continuously flow to the platform.  With a small amount of engineering effort, \tool can provide an equivalent dashboard on its mobile app, which can be used by users for viewing and controlling device states. \emph{End-to-end encryption} can be used to protect data transmissions between the local \tool mediator and the mobile app, such that data will not be disclosed to the platform or third parties (e.g., cloud-based relay servers). 
}


\vspace{3pt}\noindent\textbf{Generalizability.}
Although our prototype implementations are on SmartThings and openHAB, the presented approach can be adapted to other IoT systems. As discussed in Section~\ref{section_relay}, it is practical to implement such a mediator in most systems. On one hand, the mediator could be extended to connect a large portion of IoT devices (see Section~\ref{section_device_connector}). 
On the other hand, the mediator could interface with many platforms via a communication technology supported by the platform. Finally, a spectrum of techniques such as code analysis, UI parsing, and NLP have been employed for extracting automation rules from IoT apps or web/mobile interfaces. Therefore, \tool can be extended to support more devices and platforms with some engineering efforts. \mr{The impact of unsupported devices (which have no APIs) depends on how many such devices are deployed in a particular home. A large-scale investigation on real households that have IoT devices will be helpful to answer this question. This will be our future work.}

\vspace{3pt}\noindent\textbf{Attack surface.}
Like many existing IoT security systems (ContexIoT~\cite{jia2017contexiot}, IoTGuard~\cite{celikiotguard}, FACT~\cite{lee2017fact}),
our prototype adds an additional component to a smart home, which might become a potential target of attackers. We argue that \tool can be considered as an IoT hub (with privacy protection functionalities), so conceptually it does not make a
smart home more exploitable. \tool employs existing communication technologies used by IoT devices and platforms to connect them and does not introduce new protocols. 
Furthermore, \tool knows who it should talk with, so it
can maintain a whitelist, using an embedded firewall, to discard any incoming traffic initiated by unknown sources. 
\mr{Although it is possible that the platform could detect \tool and behave adversely, it cannot obtain more sensitive data by doing so because \tool by design only passes data that enables the execution of automation rules that are specified by users.}

\section{Conclusion}
We presented \tool, a semantics-aware customizable data flow control system
for protecting privacy of smart home owners. \tool filters data by enforcing 
data-minimization policies automatically derived from
automation applications and, optionally, user-specified policies. We overcame multiple challenges and designed an elegant virtualization-based mediating system, which enforces the policies without modifying IoT platforms or devices. 
We implemented a prototype of \tool and evaluated it in four real-world testbeds with various IoT devices of different communication protocols (ZigBee, Z-Wave, and WiFi). The evaluation results demonstrated that \tool can significantly reduce sensitive data leakage without affecting home automation. It severely impairs the attacker's ability to monitor and infer a homeowner's privacy-sensitive behaviors. Besides smart homes, \tool can also significantly enhance privacy for many other smart environments, such as smart offices, hospitals and factories.

\section*{Acknowledgment}
\fv{We thank the anonymous reviewers for their constructive suggestions. This work was supported in part by the US National Science Foundation (NSF) under grants CNS-1828363, CNS-1564128, CNS-1824440, CNS-2016589, CNS-1856380, CNS-2016415, CNS-1850278, CNS-1815144, and CNS-1953073.}

\bibliographystyle{IEEEtran}
\bibliography{main}

\appendix

\begin{figure*}[t]
  \centering
  \subfigure[]{
    \label{subfig_survey_devices}
    \includegraphics[width=0.23\textwidth]{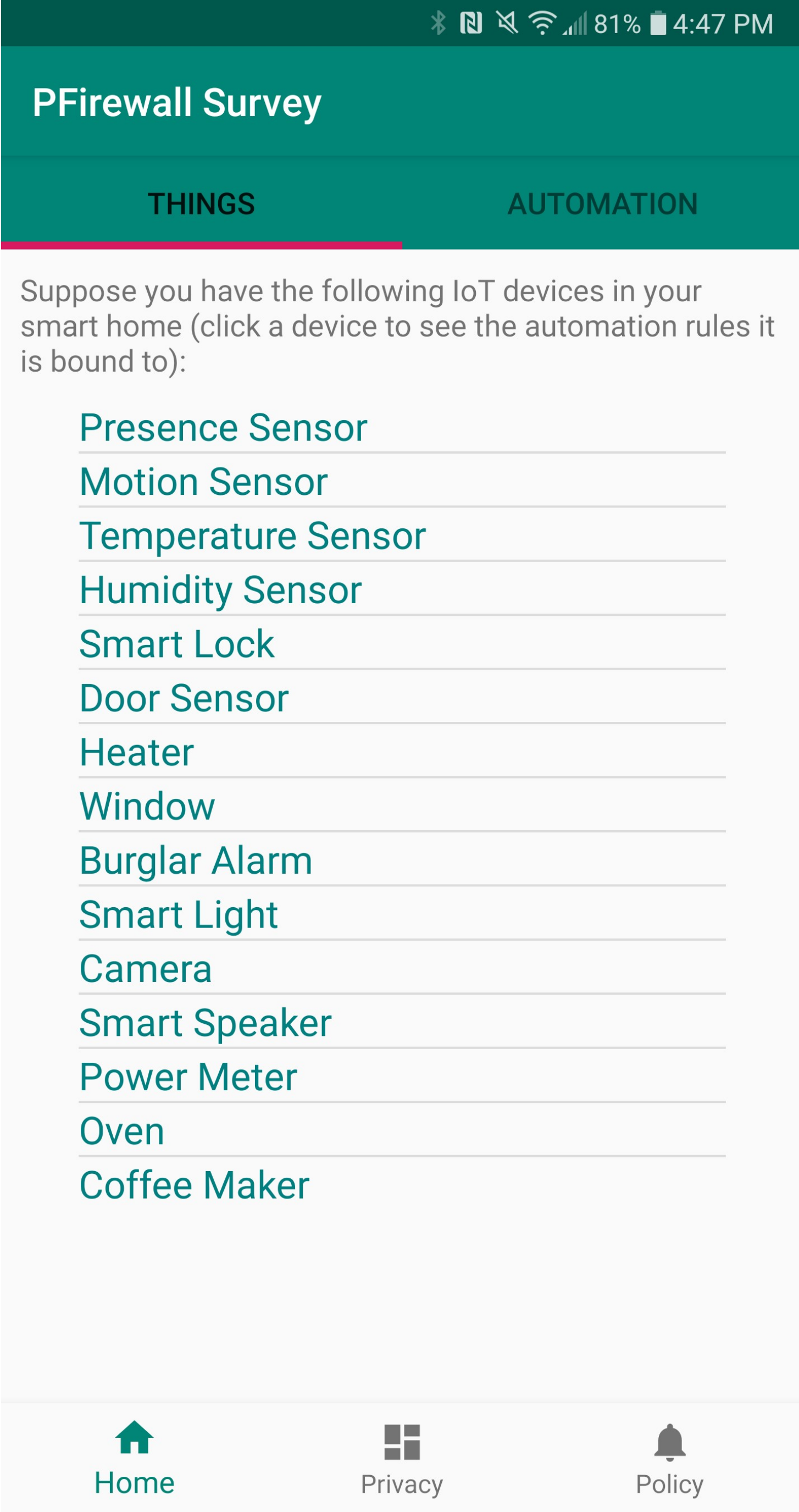}}
  \subfigure[]{
  \label{subfig_survey_automation}
     \includegraphics[width=0.23\textwidth]{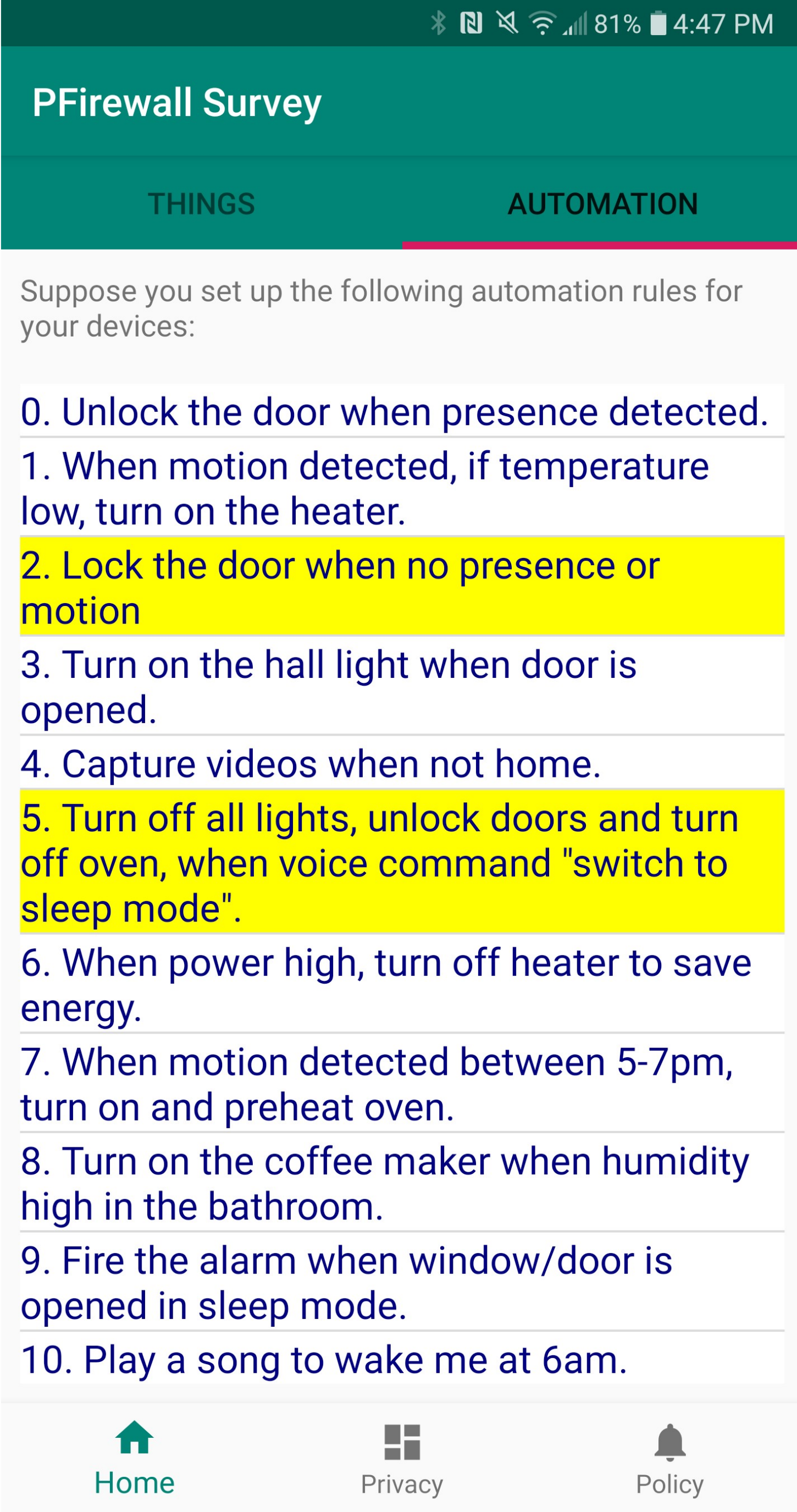}}
     \subfigure[]{
     \label{subfig_survey_tutorial}
    \includegraphics[width=0.23\textwidth]{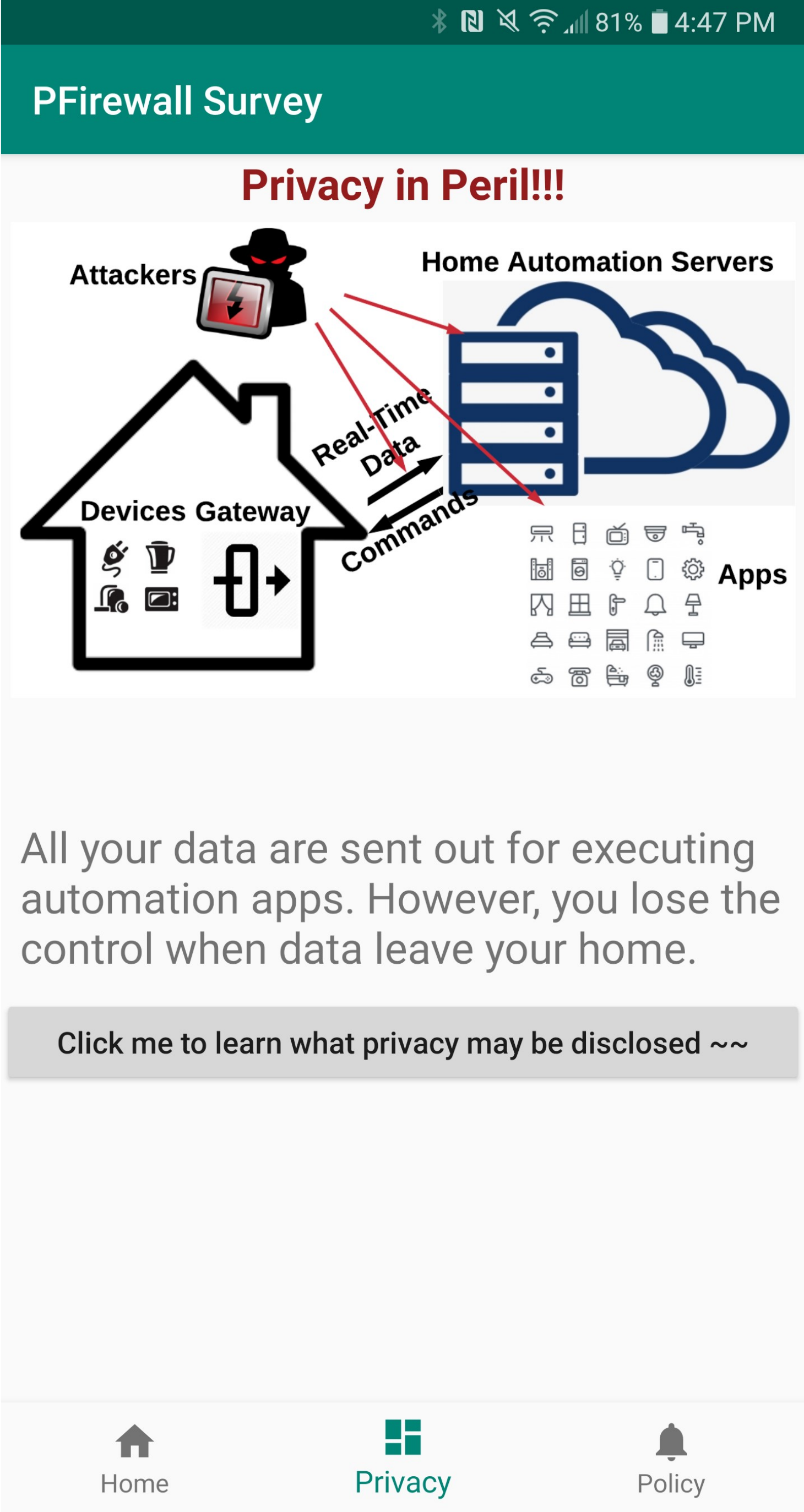}}
  \subfigure[]{
  \label{subfig_survey_privacy}
     \includegraphics[width=0.23\textwidth]{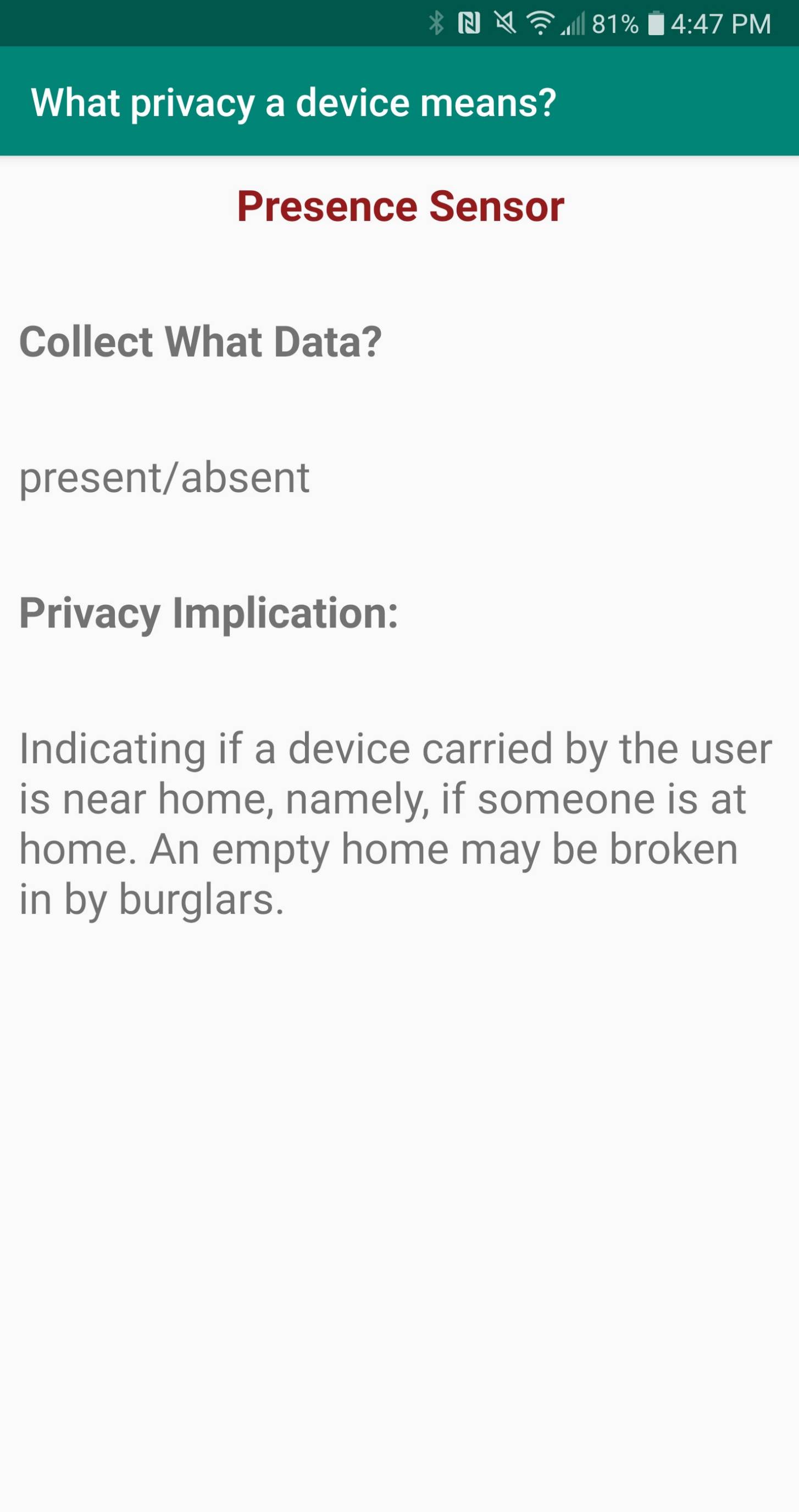}}
  \label{fig:survey}
  \caption{The PFirewall Survey mobile app used in the user survey.}
\end{figure*}

\subsection{User Study}
\label{appendix_user_study}
\subsubsection{Setup}
We conduct a user survey to study users' attitude and abilities towards defining customized data-protection policies with our policy templates (Section~\ref{section_up}). We recruit 20 adult participants who are knowledgeable about the concepts ``home automation'', ``smart home'' or ``IoT'' from our institutions. Participants completed the trial tasks of our ``PFirewall Survey'' app in our lab using smartphones we provided and after that answered several questions. 

We asked the participants to get familiar with a smart home setting where 10 automation rules (Fig.~\ref{subfig_survey_automation}) are configured to work with 15 devices (Fig.~\ref{subfig_survey_devices}). The app provides a page (Fig.~\ref{subfig_survey_tutorial}) to illustrate the architecture of the system and the potential risks of data leakage; we did not explain the content and ask questions about this page to avoid influencing the understanding of end-users by factors other than the interface itself. Besides, the app also provides an interface showing the list of 15 devices; when a device is selected, the app switches to a device detail page (e.g., Fig.~\ref{subfig_survey_privacy}) showing what data the device generates and what privacy risks are imposed if the data are leaked. In addition, policy templates (as shown in Fig.~\ref{fig_policy}) were provided for participants to define their own policies. After a 30-minute trial, participants were asked to answer questions.

\subsubsection{Results}
All 20 participants cared about their data privacy and thought it useful to define their own data flow policies for protecting privacy. However, 2 participants thought they would not spend time in defining policies even if an app is available. We collect the number of participants who had privacy concerns on each listed device. Cameras and smart speakers were the top two devices whose data are considered sensitive by the participants (19 and 16, respectively); half or more participants had concerns on the status data of smart locks, doors and windows (11, 13, 10, respectively); Each of humidity sensors, heaters, lights, powers and coffee makers is concerned by less than 3 participants. 

Besides the listed devices, the participants also cared about the data privacy of smart TVs, smart window blinds, smart outlets. 
Regarding the usability of our policy templates, 8 participants thought the templates are ``very easy'' to use and 12 participants thought them ``easy'' to use. Three participants found that they could not specify policies to control data by specifying multiple conditions with the templates, for example, the combination of a device state and a specified time period. According to the feedback, we address this issue by allowing users to choose if they would like to add a new one recursively after they complete a condition.

Overall, participants concern data privacy and hold a positive attitude in defining own policies with our templates. The result also shows that participants may overlook the privacy risks of some devices like humidity sensor and powers, which we have discussed in Section~\ref{privacy_gain}. Hence, data-minimization policies and user-specified policies could work together to achieve better privacy protection.

\subsection{IRB Approval}
\label{irb}
Our testbeds need to collect data from the testbed providers, including the 5 office members and 5 apartment members. Also, our user study involves 20 participants. We have received the approval from the IRB in the institution where all the above investigations are performed. \mr{The testbed providers and survey participants (undergraduate and graduate students) were recruited through emails and flyers. \$500 was paid to the participants of each testbed and \$50 was paid to each survey participant.}

We value the participants' privacy during our investigation processes. The data collected from all testbeds do not contain personally identifiable information and location data. The collected data will be transmitted to and stored in the password protected computer of one of the authors. Computers that store data have password-protected accounts and will be in a locked office that has limited access. Only the researchers identified on this protocol will have access to the data. Survey participants are asked to submit their questionnaire anonymously without revealing any personally identifiable information. The questionnaire will be stored in the locked office after analyses. 


\end{document}